\documentclass[preprint,numbers,12pt]{elsarticle}

\usepackage[T1]{fontenc}
\usepackage{lmodern}     
\usepackage{microtype}

\usepackage{amsmath,amssymb}

\usepackage[justification=centering]{caption}
\usepackage{subcaption}  
\usepackage{tabularx}
\usepackage{booktabs}
\usepackage{array}
\usepackage{makecell}

\usepackage{algorithm}
\usepackage{algorithmic} 
\usepackage{lineno}      
\usepackage{lipsum}

\usepackage{hyperref}
\hypersetup{
    colorlinks=true,
    allcolors=blue,
}
\usepackage{url}

\journal{Neurocomputing}

\begin{document}

\begin{frontmatter}



\title{GCMCG\@: A Clustering-Aware Graph Attention and Expert Fusion Network for Multi-Paradigm, Multi-task, and Cross-Subject EEG Decoding}


%\author{} %% Author name
\author[labela,equal]{Yiqiao Chen}
\author[labela,equal]{Zijian Huang}
\author[labelb,labeld,equal]{Juchi He}
\author[labelb,labelc,equal]{Fazheng Xu}
\author[labela]{Zhenghui Feng\corref{cor1}}

%\affiliation{organization={},%Department and Organization
\affiliation[labela]{
    organization={Faculty of frontier sciences, Harbin Institute of Technology, Shenzhen}, 
    addressline={Harbin Institute of Technology Shenzhen Campus, University Town of Shenzhen},
    city={Shenzhen},
    postcode={518055},
    state={Guangdong Province},
    country={China}}

\affiliation[labelb]{
    organization={Doumaomao NetTech Co., Ltd.}, city={Luoyang}, country={China}}

\affiliation[labelc]{
    organization={University of Melbourne},country={Australia}}

\affiliation[labeld]{
    organization={The University of New South Wales},country={Australia}
}

\cortext[cor1]{Corresponding author at: fengzhenghui@hit.edu.cn}
\fntext[equal]{The four authors contribute equally to this work.}

%% Abstract
\begin{abstract}
Brain-Computer Interfaces (BCIs) based on Motor Execution (ME) and Motor Imagery (MI) electroencephalogram (EEG) signals offer a direct pathway for human-machine interaction. However, developing robust decoding models remains challenging due to the complex spatio-temporal dynamics of EEG, its low signal-to-noise ratio, and the limited generalizability of many existing approaches across subjects and paradigms. To address these issues, this paper proposes \textbf{G}raph-guided 
\textbf{C}lustering \textbf{M}ixture-of-Experts 
\textbf{C}NN-\textbf{G}RU (GCMCG), a novel unified framework for MI-ME EEG decoding. Our approach integrates a robust preprocessing stage using Independent Component Analysis and Wavelet Transform~(ICA-WT) for effective denoising. We further introduce a pre-trainable graph tokenization module that dynamically models electrode relationships via a Graph Attention Network (GAT), followed by unsupervised spectral clustering to decompose signals into interpretable functional brain regions. Each region is processed by a dedicated CNN-GRU expert network, and a gated fusion mechanism with L1 regularization adaptively combines these local features with a global expert. This Mixture-of-Experts (MoE) design enables deep spatio-temporal fusion and enhances representational capacity. A three-stage training strategy incorporating focal loss and progressive sampling is employed to improve cross-subject generalization and handle class imbalance. Evaluated on three public datasets of varying complexity (EEGmmidb-BCI2000, BCI-IV 2a, and M3CV), GCMCG achieving overall accuracies of $86.60\%$, $98.57\%$, and $99.61\%$, respectively, which demonstrates its superior effectiveness and strong generalization capability for practical BCI applications.

\end{abstract}



%% Keywords

\end{frontmatter}



\section{Introduction}\label{sec:Introduction}
%% Labels are used to cross-reference an item using \ref command.
The Brain-Computer Interface~(BCI), 
a cutting-edge Human-Machine Interaction~(HMI) technology,
enables communication between the brain and external devices by decoding thoughts from neural signals, 
effectively bypassing traditional muscular pathways~\cite{mudgal2020brain}.
Among various neuroimaging modalities, electroencephalogram~(EEG) based BCI systems	have emerged as a rapidly advancing and widely adopted approach due to their cost-effectiveness, non-invasiveness,
and user-friendliness. 

The mainstream paradigms of EEG-based BCIs include Motor Execution (ME) and Motor Imagery (MI), both of which focus on decoding motor-related brain activity. Distinct from paradigms that rely on external triggers to evoke brain responses, ME and MI capture neural signals associated with motor intentions without such external inputs. ME entails actual physical movement, whereas MI involves only the mental rehearsal of actions without muscle activation~\cite{pan2024comprehensive}. Despite their differences, both paradigms elicit comparable EEG patterns, such as event-related desynchronization (ERD) and synchronization (ERS). These ERD/ERS patterns serve as the physiological basis for translating both executed and imagined motor intentions into control commands in BCI systems, enabling the decoding of tasks such as hand or foot movements regardless of whether they are physically performed or mentally simulated. Given these shared mechanisms, it is theoretically feasible to integrate data from both ME and MI paradigms to expand sample sizes and enhance model training.
% The mainstream paradigms of EEG-based BCIs include Motor Execution (ME) and Motor Imagery (MI), which focus on decoding motor-related brain activity. Distinct from paradigms that rely on external triggers to evoke brain responses, ME and MI capture neural signals associated with motor intentions without such external inputs. 
% ME entails actual physical movement, 
% whereas MI involves only the mental rehearsal of actions~\cite{meng2018differences,tayeb2019validating}, 
% without muscle activation—yet both paradigms elicit comparable EEG patterns, 
% such as event-related desynchronization (ERD) and synchronization (ERS).
% These ERD/ERS patterns serve as the physiological basis for translating both executed and imagined motor intentions into control commands in BCI systems~\cite{zhang2019using,khademi2023review}, 
% enabling the decoding of tasks such as hand or foot movements regardless of whether they are physically performed or mentally simulated.
Consequently, researchers in the BCI field often utilize data from ME/MI paradigms to discriminate human motor behaviors, employing a wide variety of machine learning (ML) and deep learning (DL) models and architectures.
%Numerous studies have investigated Machine Learning (ML) and Deep Learning (DL) approaches for classifying hand or foot movements (both executed and imagined) using EEG-based ME and MI paradigms.

%In the process of signal processing and data augmentation, previous works ... 

Early works applied techniques such as 
Support Vector Machine~(SVM), 
and Filter Bank Common Spatial Pattern~(FBCSP)~\cite{mebarkia2019multiSVM,ang2012FBCSP} , 
%我们做多分类ML做不了
establishing a foundational pathway for EEG decoding through basic preprocessing, manual feature extraction, and classical classifiers. 
Building upon these, 
recent deep learning models—including Convolutional Neural Network (CNN)-based 
methods~\cite{lawhern2018eegnet, LIU2023CMOCNN}, 
Recurrent Neural Network (RNN) variants~\cite{luo2018rnnGRU, zhang2017intent,karimian2024tdlstm}, 
and Transformer architectures~\cite{song2021transformer}—have demonstrated strong capabilities 
in learning hierarchical, data-driven representations directly from raw EEG signals. 
By designing diverse and complex network structures, these models have shown improvements in capturing spatio-temporal patterns, enhancing classification accuracy, and facilitating cross-subject generalization.
%These models have shown improvements in capturing complex spatial-temporal patterns and in enhancing classification accuracy and cross-subject generalization. 

Nevertheless, each of these architectures has inherent limitations stemming from their structural designs: CNNs exhibit limited capacity for modeling long-range temporal dependencies, 
RNNs often suffer from vanishing gradients and training instability~\cite{karimian2024tdlstm}, 
and Transformer-based approaches typically require large-scale data to perform reliably~\cite{wang2023largeTF, chen2025LCMpretrained}.
To further enhance model performance, 
composite deep architectures have been explored, FFCL
including CNN-LSTM hybrids~\cite{li2022FFCL} that combine spatial feature extraction with temporal modeling; 
GLT-Net
adapts Graph-based Neural Network~(GNN)~\cite{aung2025eeg_gltGCN} that explicitly model relationships between electrodes.
However, many existing spatio-temporal approaches simply concatenate spatial and temporal features just before the final fully connected layers for prediction. This late and mechanical fusion fails to facilitate deeper, synergistic interaction between spatial and temporal representations within the network. Consequently, although such architectures generally outperform models that focus solely on either spatial or temporal features, their performance remains suboptimal, indicating significant room for improving the fusion of complementary information streams.
%while these architectures surpass models focusing solely on either spatial or temporal features, their performance remains suboptimal, indicating significant room for improvement in effectively fusing these complementary information streams.

In addition, some existing works are evaluated primarily 
in within-subject settings~\cite{luo2018rnnGRU,aung2025eeg_gltGCN}, 
where training and testing are conducted on data from the same individual. 
While useful for benchmarking, 
such setups may not fully reflect the demands of cross-subject generalization required in realistic deployments. 
Some studies also conflate intra- and cross-subject settings 
by fusing discrete samples or using single-dataset validation~\cite{karimian2024tdlstm}.

Furthermore, while several deep learning methods operate directly on raw EEG signals~\cite{LIU2023CMOCNN,lawhern2018eegnet, karimian2024tdlstm,song2021transformer}, they often overlook the impact of inherent noise on classification accuracy. EEG signals are characterized by a low signal-to-noise ratio (SNR), non-stationarity, and susceptibility to various physiological artifacts (e.g., from ECG, EOG, and EMG), which complicate the extraction of discriminative features. Effective denoising is therefore critical for models to reliably capture task-relevant information from EEG data.
%EEG signals are inherently noisy, with low signal-to-noise ratio~(SNR), variable sequence lengths, and heterogeneous electrode configurations. EEG signals are notoriously difficult to denoise primarily due to contamination from multiple sources, especially physiological artifacts such as ECG (heartbeats), EOG (eye movements and blinks), and EMG (muscle activity). These interferences often overlap with EEG signals in the frequency domain, making them hard to isolate or remove. Under such complex noise conditions, EEG signals exhibit strong nonlinearity, nonstationarity, and non-Gaussianity, which challenge conventional denoising methods based on simple mathematical assumptions. Moreover, their low amplitude makes them particularly vulnerable to being overwhelmed by noise in the time domain.Taken together, these observations highlight the continued need for decoding frameworks that are more flexible and capable of accommodating diverse paradigms, realistic evaluation settings, and signal variability in practical BCI applications.

Taken together, these observations underscore the continued need for decoding frameworks that are more flexible in capturing and integrating spatio-temporal dynamics, robust to noise, and capable of performing well under realistic evaluation settings with diverse subjects and paradigms.

\par
In this paper, 
we propose \textbf{GCMCG}, 
a unified and robust framework for decoding MI-ME EEG signals. The framework integrates several novel components to enhance interpretability, generalization, and performance.
First, raw EEG signals are preprocessed using Independent Component Analysis and Wavelet Transform~(ICA-WT) to suppress physiological and environmental noise, significantly improving signal quality and robustness.
%First, ICA and Wavelet Transform are used to suppress physiological and environmental noise, improving robustness and signal quality. 
Second, we introduce a pretrainable tokenization module that encodes electrode spatial topography into a learnable graph structure, treating each electrode as a node and enabling dynamic, functionally-aligned graph construction. This graph is processed by a Graph Attention Network (GAT) to adaptively capture spatiotemporal relationships without directly hand-defined adjacency, followed by unsupervised spectral clustering that decomposes the graph into interpretable brain-region segments.
%To enhance cross-subject generalization and electrode adaptability, we introduce a pretrainable tokenization module that encodes spatial topography into a learnable graph structure. Each electrode is treated as a node, enabling dynamic graph construction aligned with functional brain regions.Second, a Graph Attention Network~(GAT) is used to adaptively learn temporal dependencies and spatial relationships, without requiring manually defined adjacency.  
Third, each functional region is assigned a dedicated CNN-GRU expert to model localized spatiotemporal patterns, while a gated fusion mechanism with L1 regularization and learnable attention masks adaptively combines outputs from both local and global experts. This mixture-of-experts (MoE) design promotes specialization, reduces overfitting, and yields richer, more interpretable representations.
%Building on this graph representation,unsupervised spectral clustering segments the graph into interpretable functional regions, supporting biologically inspired spatial decomposition.Each region is assigned to a CNN-GRU expert for localized spatiotemporal decoding, and expert outputs are adaptively fused via a gated mechanism with L1 regularization and learnable attention masks. By integrating both global and region-specific GRU experts, GCMCG effectively captures holistic brain dynamics and localized functional dependencies, enabling richer and more interpretable spatiotemporal representations.This MoE modular fusion mitigates overfitting and enhances specialization. 
Finally, to accommodate real-world variability, we implement a three-stage training strategy incorporating focal loss, progressive sampling, and learnable weight scaling, effectively tackling class imbalance and enhancing cross-subject generalization. Moreover, through dynamic masking and graph-adaptive design, GCMCG natively supports variable-length sequences and diverse electrode layouts, narrowing the gap between deep learning models and practical BCI deployment.
%Beyond architectural design, we further refine the framework with practical training strategies and data handling mechanisms.To accommodate class imbalance and multi-class MI-ME decoding, we further introduce a three-stage training scheme combining focal loss, progressive sampling, and learnable weight scaling to improve training stability and generalization.Moreover, GCMCG naturally supports variable-length sequences and diverse electrode layouts through its dynamic masking and graph-adaptive design, thereby bridging the gap between practical BCI deployment and traditional deep learning models.

\par Leveraging the strengths of the GCMCG framework—including enhanced signal denoising, deep spatiotemporal feature extraction and fusion, and a robust training scheme—we evaluated our model on three public EEG datasets with varying classification complexities: 
EEGmmidb BCI2000 (9-class), BCI-IV 2a (4-class), and M3CV (3-class). The proposed model achieves overall accuracies of $86.60\%$, $98.57\%$, and $99.61\%$, respectively. These results highlight the effectiveness of GCMCG in handling diverse task complexities and electrode configurations, underscoring its generalization capability across different experimental settings.
%We evaluated GCMCG on four public EEG datasets with varying classification complexities: BCI-IV 2b (2 classes), BCI-IV 2a (4 classes), EEGmmidb BCI2000 (9 classes), and M3CV (6 classes).The proposed model achieves  the state-of-the-art average accuracies of 99.29\%, 95.09\%, 89.29\%, and 92.83\%, with corresponding kappa values of XX.59\%, XX.19\%, 88.59\%, and XX.67\%, respectively.

\par
%Our main contributions are the following:
Benefiting from the aforementioned design, GCMCG demonstrates superior performance across multiple benchmarks. To summarize, this work makes the following key contributions:
\begin{itemize}
    \item A robust denoising and adaptive input pipeline that mitigates EEG noise via ICA-WT, while flexibly supporting variable-length sequences and heterogeneous electrode configurations.
    \item A unified spatio-temporal modeling framework that deeply integrates graph attention networks with spectral clustering and gated multi-expert fusion, enabling synergistic feature learning.
    \item A stable and generalizable training strategy that addresses class imbalance and variability through focal loss, progressive sampling, and learnable scaling, enhancing cross-subject robustness.
\end{itemize}
The remainder of this paper is organized as follows: 
Section~\ref{sec:RelatedWorks} reviews related studies. 
Section~\ref{sec:Method} describes the GCMCG framework in detail. 
Section~\ref{sec:Experiments} presents the experimental settings and results. 
Section~\ref{sec:Discussion} points some findings when manipulating EEG classification tasks.
Section~\ref{sec:Conclusion} concludes the paper and outlines future work.

\section{Related Works}\label{sec:RelatedWorks}

This section reviews representative methods for EEG-based motor task classification to contextualize the  contributions of our GCMCG framework. Although our approach introduces integrated innovations in signal denoising, model architecture, and training strategy, the field's mainstream research and comparative benchmarks are predominantly centered on model architectures. Therefore, this review follows this convention, tracing the evolution from classical machine learning pipelines reliant on manual feature engineering to modern deep learning paradigms. We critically examine convolutional, recurrent, and graph-based architectures, with a specific focus on how they model and fuse spatio-temporal dynamics. This structured overview delineates the strengths and limitations of existing approaches, thereby clarifying the specific architectural gaps that our framework addresses, particularly the need for deeper, more synergistic spatio-temporal integration.

% To support the construction of our framework, 
% we conduct a structured review of representative methods 
% for EEG-based motor task classification.  
% The review includes both machine learning pipelines 
% based on manual features and deep learning approaches, including convolutional, 
% recurrent, and graph-based models.  
% Special attention is paid to multi-branch architectures, 
% spatial-temporal modeling, and the role of feature fusion.  
% These categories reflect the core components integrated 
% in our proposed GCMCG framework and help to contextualize its architectural choices 
% and contributions.

\subsection{Machine Learning}
%类别1：特征提取+传统机器学习

% ML methods are approaches typically involving signal 
% preprocessing, feature extraction, and classification. 
% Preprocessing methods such as filtering~\cite{blankertz2007optimizingFilters_CSP} or 
% Independent Component Analysis~(ICA)~\cite{lu2006ica,Kachenoura2008ICA_BCI} aims to improve signal 
% quality but may also distort relevant information.

% For feature extraction, 
% classical methods have made meaningful contributions, 
% particularly in enhancing EEG discriminability through manual representations. 
% Common Spatial Pattern (CSP)\cite{blankertz2007optimizingFilters_CSP,Gaur2021CSP_EEGBCI} 
% and its extensions like Filter Bank CSP (FBCSP)\cite{ang2012FBCSP} 
% are widely adopted to optimize spatial filters. 
% Meanwhile, time-frequency approaches—such as 
% Continuous Wavelet Transform (CWT)\cite{lee2019applicationCWT} 
% and Short-Time Fourier Transform (STFT)\cite{tabar2016novelSTFT}—have been used to 
% capture non-stationary temporal dynamics. These feature engineering techniques remain influential, 
% offering valuable insights for later neural network designs, 
% particularly in how EEG representations are decomposed and preprocessed.

ML methods for EEG decoding often rely on manual feature extraction, as their performance is typically constrained by the dimensionality and complexity of raw neural signals. These approaches generally follow a multi-stage pipeline involving signal preprocessing, feature engineering, and classifier training. For preprocessing, techniques such as filtering~\cite{blankertz2007optimizingFilters_CSP} or Independent Component Analysis (ICA)~\cite{lu2006ica,Kachenoura2008ICA_BCI} are commonly employed to attenuate noise and enhance signal quality, yet they may inadvertently distort or remove physiologically relevant information. Because filtering can suppress or smear signal components that fall outside predefined passbands or overlap with noise in the frequency domain, while ICA risks removing neurally-generated activity if the statistical independence assumption is not perfectly met or if artifact and brain signal components exhibit overlap. 

For feature extraction, classical methods have made meaningful contributions by constructing discriminative representations from preprocessed EEG. Notable among these is Common Spatial Pattern (CSP)~\cite{blankertz2007optimizingFilters_CSP,Gaur2021CSP_EEGBCI} and its extensions such as Filter Bank CSP (FBCSP)~\cite{ang2012FBCSP}, which learn spatially selective filters to maximize inter-class variance. Complementing these, time-frequency representations, including Continuous Wavelet Transform (CWT)~\cite{lee2019applicationCWT} and Short-Time Fourier Transform (STFT)~\cite{tabar2016novelSTFT}, have been widely adopted to characterize non-stationary dynamics in motor-related EEG. While these manually designed features continue to inform later deep learning designs, particularly in guiding representation learning and preprocessing strategies, they inherently limit model capacity in capturing complex spatio-temporal dependencies in an end-to-end manner.

% For classification, traditional models like LDA, SVM, PCA, 
% and DT~\cite{narayan2021LDA,ZHOU2021PCA,mebarkia2019multiSVM} 
% have provided early baselines, 
% but their modeling capabilities are relatively shallow. 
% These models often rely on linear assumptions or static decision boundaries, 
% making it difficult to capture nonlinear or long-range dependencies in EEG sequences. 
% Their generalization performance, 
% especially under cross-subject or multi-class MI settings, remains limited.
For classification, ML models are fundamentally limited by their structural assumptions, preventing them from achieving high decoding accuracy on complex EEG data. Methods such as Linear Discriminant Analysis (LDA) and Support Vector Machines (SVM) construct linear or shallow decision boundaries, while Principal Component Analysis (PCA) performs linear dimensionality reduction, and Decision Trees (DT) employ axis-aligned partitions of the feature space~\cite{narayan2021LDA,ZHOU2021PCA,mebarkia2019multiSVM}. These structural characteristics restrict their capacity to model the nonlinear and dynamic spatio-temporal patterns inherent in motor-related EEG. Consequently, their generalization capability remains limited, particularly in challenging scenarios involving cross-subject variability or multi-class motor imagery tasks.

% The rise of deep learning offers a natural evolution in this context. 
% By building on the signal representations shaped by classical methods, 
% modern architectures such as CNNs, RNNs, and attention-based models 
% have introduced more expressive mechanisms to learn hierarchical 
% and latent relationships between EEG signals and target variables. 
% In particular, many preprocessing strategies (e.g., frequency filtering, CWT) 
% originally developed in the ML era continue to serve as integral components in DL pipelines, 
% highlighting a lineage of ideas rather than a strict methodological divide.

\subsection{Deep Learning}

%As computing resources have become more powerful and accessible,
%deep learning~(DL) has garnered increasing attention in recent years. 
%DL methods are capable of 
%automatically learning hierarchical and latent features 
%directly~\cite{altaheri2023deep}.
% DL approaches have gained increasing attention for their ability to model complex, 
% nonlinear relationships between input signals and target variables. 
% Rather than relying on manual features, 
% DL methods learn to represent hierarchical 
% and latent structures directly from data~\cite{altaheri2023deep}, 
% enabling more flexible and scalable modeling. 
% Notably, many modern architectures inherit core ideas from classical pipelines—such as 
% time-frequency decomposition and spatial filtering—while embedding them into trainable, 
% end-to-end frameworks that can better adapt to subject variability and signal complexity.

DL models constitute a natural evolution from ML models, addressing its core limitations through end-to-end representation learning. For preprocessing, while some DL studies directly employ raw EEG signals and report competitive performance, these results do not inherently negate the value of preprocessing. Rather, they reveal the constraints of certain ML-era methods, such as basic filtering or fixed time-frequency transforms, which remain relatively common in DL pipelines. Designed to simplify data for manual analysis, these techniques can inadvertently discard valuable information, creating an unnecessary bottleneck for the model.

For model structure, a key strength of DL lies in its ability to automatically learn hierarchical and discriminative features directly from data, without relying on handcrafted feature engineering~\cite{altaheri2023deep}. Architectures such as CNNs, RNNs, and attention-based models have introduced more expressive mechanisms for capturing complex, nonlinear relationships in EEG signals. Importantly, these models often embed core ideas from classical methods into flexible neural modules, such as time-frequency analysis and spatial filtering, which enable end-to-end optimization while maintaining interpretability and physiological plausibility.

% from raw EEG data

%%% 应该先讲GRU+GNN，然后， Comparison of the performance of LSTM
%% and CNNs compared to time series data
%% 然后2.3. Multi-branch network and feature fusion 
%% 或者cnn先讲，因为也确实用到了cnn，但要简单提cnn方法是什么

%%可能需要提及一些工作的结果

%\subsubsection{CNN} 
\subsubsection{Modeling Local Spatial Patterns}
 
To enhance feature representations, studies preprocess raw EEG signals
with time-frequency transformations. 
Wang et al.~\cite{wang2022mi} applied 
Shannon complex wavelet transform before feeding 
the data into a modified ResNet. 
Similarly, Krishnan et al.~\cite{keerthi2021cnn} 
used variational mode decomposition and 
short-time Fourier transform~(STFT) to generate EEG spectrograms
for improved CNN-based classification performance.
However, such preprocessing may lose important high-frequency 
and inter-channel information, because these techniques are governed by the time-frequency uncertainty principle and, more critically, these methods are typically applied to each EEG channel in isolation. To address this, recent research has explored end-to-end CNN architectures 
that process raw EEG signals directly and incorporate multi-level or multi-branch feature fusion. 
Amin et al.~\cite{amin2019multilevel} proposed a multilevel weighted CNN 
that extracts features at different depths and fuses them via fully connected layers, 
enhancing decoding capacity. Nevertheless, the architectural complexity introduced by such multi-branch schemes presents a significant drawback. This impasse underscores the need for a new denoising pathway within the EEG deep learning framework that can effectively suppress artifacts without discarding salient neural information or incurring prohibitive computational overhead.

In terms of model architecture, CNNs have been widely adopted for EEG-based classification due to their strong capability in automatically extracting spatial and local features~\cite{rajwal2023convolutional,yang2024novel}. Among them, EEGNet~\cite{lawhern2018eegnet} has emerged as a widely used baseline, employing depthwise separable and temporal convolutions to efficiently learn spatial-frequency representations. Further advancing this line, Liu et al.~\cite{LIU2023CMOCNN} proposed CMO-CNN—a compact multi-branch 1D CNN architecture incorporating branches of varying depths, residual blocks, and 1D squeeze-and-excitation (SE) modules. This design enables the model to learn rich multi-scale representations while maintaining parameter efficiency. Evaluated on raw EEG inputs, CMO-CNN demonstrates strong performance in both within-subject and cross-subject settings. Under Leave-One-Subject-Out (LOSO) cross-validation, it achieved average accuracies of $83.92\%$ on a four-class task and $87.19\%$ on a binary task. Despite their proficiency in capturing local spatial features and their successful application in EEG decoding, CNNs exhibit limited capacity in modeling long-range temporal dependencies and adapting to the inherent variability of EEG data, such as noise and inter-subject differences. These limitations have motivated the integration of complementary techniques, such as temporal modeling mechanisms and graph-based modules, in subsequent research.
\subsubsection{Capturing Temporal Dependencies}

Modeling temporal dependencies is essential in EEG decoding, and RNNs offer a natural architectural choice for this purpose. However, traditional RNNs are limited by vanishing gradients, which hinder their ability to capture long-range dependencies. To address these issues, advanced variants such as LSTM and GRU have been developed.
% Modeling temporal dependencies is essential in EEG decoding, 
% and RNNs are a natural fit for this purpose. 
% For instance, the CasRNN architecture~\cite{zhang2018cascade} 
% integrates convolutional and recurrent layers for 
% EEG motor intention recognition.
% However, traditional RNNs suffer from vanishing gradients, 
% which limit their performance on long sequences. 
% To overcome these limitations, 
% advanced variants such as LSTM and GRU have been proposed.

LSTM employs memory cells with three gating mechanisms to preserve information over extended time steps, making it suitable for capturing the long-range temporal structure of EEG signals. Its effectiveness in combination with signal decomposition techniques has been well demonstrated.
For example, TD-LSTM~\cite{karimian2024tdlstm}, adopted time-distributed blocks to enhance per-timestep feature representation and capture fine-grained temporal dependencies, achieving state-of-the-art results in joint motor imagery–motor execution classification. Nevertheless, in EEG decoding, the increased complexity and parameter count of LSTM-based models do not consistently yield performance gains—a point we further explore in our experiments.
% LSTM uses memory cells with three gating mechanisms to retain information over long time steps, suiting EEG signals’ long-range temporal structure. Its effectiveness with signal decomposition is well-demonstrated: Li et al.~\cite{li2016combined} proposed a DWT-LSTM for time-frequency analysis, Lin et al.~\cite{lin2020bci} introduced a DWT-BiLSTM for bidirectional learning, and Mwata et al.~\cite{mwata2022empirical} used empirical mode decomposition with BiLSTM for adaptive sub-band features. Other models~\cite{karimian2024tdlstm} adopted time-distributed LSTM blocks to boost per-timestep feature extraction and time-resolved dependencies, achieving SOTA in joint MI-ME classification. However, for EEG decoding, LSTM-based models’ increased complexity and parameters do not always improve performance—this is further examined in our experiments.

GRU offers a streamlined alternative to LSTM by combining the forget and input gates into a single update gate, thereby reducing the number of trainable parameters. This design makes GRU especially attractive in scenarios with limited training data or where faster convergence is desired~\cite{luo2018rnnGRU,narotamo2024deepECG}. In recent biomedical applications, GRU-based models have demonstrated competitive performance with lower computational overhead. For example, in ECG classification, GRUs have matched or even surpassed the generalization capability and training efficiency of LSTMs, particularly when embedded within compact CNN-GRU frameworks~\cite{narotamo2024deepECG}. These attributes position GRU as a suitable candidate for modular and lightweight EEG decoding systems, where balancing model complexity and performance is crucial~\cite{luo2018rnnGRU}. 

% GRU~\cite{dey2017gateGRU,cho2014learningGRU}, is a simplified alternative to LSTM that merges the forget 
% and input gates into a single update gate, 
% effectively reducing the number of trainable parameters. 
% This makes GRU particularly appealing for scenarios with limited training data 
% or requiring faster convergence~\cite{luo2018rnnGRU,narotamo2024deepECG}.
% In recent biomedical applications, 
% GRU-based models have demonstrated competitive performance 
% with lower computational cost. For example, in ECG classification tasks, 
% GRUs have achieved generalization and training efficiency comparable to 
% LSTM counterparts, and in some cases, 
% even outperformed them—
% especially when integrated into compact architectures 
% such as CNN-GRU frameworks~\cite{narotamo2024deepECG}.
% These characteristics make GRU a suitable candidate for modular, 
% lightweight EEG decoding frameworks, 
% where balancing model complexity and performance is critical~\cite{luo2018rnnGRU}. 

Although CNN-GRU architectures have achieved promising results in related biomedical domains, they often fall short in effectively integrating cross-channel dependencies inherent in multi-channel EEG signals, which is a critical aspect for capturing the coordinated activity of brain regions. In this context, the structural simplicity, parameter efficiency, and robust temporal modeling capability of GRU make it particularly well-suited for incorporation into more advanced architectures aimed at unifying spatial and temporal feature learning. Owing to these comparative advantages among sequential models, GRU represents a compelling choice for our proposed model, where both parameter efficiency and temporal generalization are critically important.

% While LSTM remains a strong baseline for modeling 
% long-range temporal dependencies, the structural simplicity 
% and efficiency of GRU offer greater potential for integration 
% into unified spatial-temporal systems.
% CNN-GRU architectures have shown promise in related domains; 
% however, they do not appear to have undergone practical evaluation 
% in the context of EEG decoding, calling for further empirical investigation.
% Overall, GRUs strike a good balance between accuracy and efficiency, and demonstrate better resistance to overfitting.
% These advantages make GRU a compelling choice in our proposed model, 
% where parameter efficiency and temporal generalization are both critical.

%\subsubsection{GAT}
\subsubsection{Integrating Inter-Channel Relationships}
%类别3：图神经网络

%EEG signals are recorded from spatially distributed scalp electrodes
Beyond the rich temporal dynamics, EEG data possesses another critical dimension: it is inherently multi-channel, recorded from an array of spatially distributed scalp electrodes. This configuration naturally lends itself to a graph representation, where each electrode is treated as a node in a graph, and the signals collected from these nodes form a sequence of data generated from this graph structure; conventional CNN and RNN models, however, face challenges in processing such data—they rely on fixed grid-like (for CNN) or sequential (for RNN) structures that do not naturally fit the graph-based nature of EEG electrode layouts. Notably, defining reliable connections between these electrode nodes (i.e., graph edges) is non-trivial: simple methods like using Euclidean distance often yield unstable results across recording sessions due to factors such as head movement, individual anatomical variability, and non-uniform electrode placement, which makes GNNs a suitable choice as they can explicitly model the spatial relationships among electrode nodes.

This perspective has motivated a growing number of studies to adopt graph-structured frameworks for more accurate inter-channel relationship modeling and, consequently, improved decoding performance.
While Shi et al.~\cite{shi2024brainGNN} embedded topography-aware graph representations 
into a CNN framework using fixed spatial priors, 
other works have explored more explicit graph modeling for EEG decoding. 
Song et al.\cite{Song2020DGCNN} introduced a dynamical graph CNN (DGCNN) 
that adaptively constructed adjacency matrices from input features, 
though limited to emotion recognition and using global graph templates. 
Aung et al.\cite{aung2025eeg_gltGCN} proposed GLT-Net, which optimizes 
a multi-stage graph learning pipeline guided by temporal-frequency correlations. 
Their approach introduces learned adjacency refinement, 
yet the resulting graph structure remains static once optimized, 
relying on predefined connectivity rather than adaptive tokenization.

% Among GNN variants, 
% GATs are particularly suited to 
% EEG modeling 
To overcome this limitation of static graph construction, the Graph Attention Network (GAT) has emerged as a promising alternative, primarily due to their ability to assign learnable, 
input-dependent attention weights to edges. This mechanism enables dynamic adaptation to signal variations across trials and subjects while preserving the structural symmetry of the underlying electrode graph. 
A few recent studies have incorporated GATs into EEG pipelines 
to enhance spatial representation learning, 
often within spatiotemporal frameworks that 
jointly extract electrode relationships and temporal dependencies. 
For example, Ma et al.\cite{ma2023mbgaGAT} proposed MBGA-Net, 
a multi-branch graph attention network that leverages subject-specific spatial graphs 
to improve individualized MI classification. 

While these efforts demonstrate the promise of GAT-based architectures, many existing approaches in EEG decoding continue to favor predefined or static graph topologies—particularly in temporal modeling frameworks capable of capturing richer signal dynamics. These methods often assume fixed input dimensionality and electrode layouts and are predominantly designed for binary motor imagery tasks. Their monolithic architectural design may further limit adaptability in more complex scenarios, such as multi-class classification, multi-paradigm decoding, or cross-subject generalization, where modular and flexible modeling is essential.
% While these efforts demonstrate the promise of GAT-based architectures, 
% many existing approaches still rely on predefined or 
% static graph structures, 
% assume fixed input dimensionality and electrode layouts, 
% and are primarily designed for binary MI tasks. 
% Their monolithic architectures may present challenges 
% in adapting to multi-class, multi-paradigm, 
% or cross-subject scenarios, where greater modularity 
% and flexibility are desirable.
% Also, directly feeding raw EEG signals into graph-based architectures 
% can be computationally demanding and less efficient, 
% especially when processing high-resolution temporal sequences.

EEG signals are characterized by non-invasive acquisition, multi-channel layout, and complex temporal dynamics. While existing studies have typically addressed individual aspects of these characteristics, such as denoising, temporal modeling, or spatial interaction, a unified framework that comprehensively integrates all essential EEG attributes remains lacking. To bridge this gap, we propose an architecture that incorporates adaptive graph learning with clustering and modular decoding, eliminating reliance on predefined adjacency constraints and fixed spatial-temporal processing pathways. This design dynamically captures inter-channel dependencies, supports multi-scale temporal feature extraction, and ensures robust generalization across diverse EEG configurations and behavioral tasks.
% These limitations highlight the need for graph-based EEG models 
% that are both structure-aware and dynamically adaptable. 
% To this end, our work adopts a GAT-based design that 
% avoids reliance on predefined adjacency matrices and 
% supports flexible modeling of inter-electrode relationships. 
% In contrast to existing frameworks that process spatial-temporal 
% information through fixed or monolithic paths, 
% our approach integrates adaptive graph learning with clustering and modular decoding, 
% enabling robust performance across heterogeneous EEG configurations and decoding tasks.

\begin{figure*}[htbp]
	\centering
	\includegraphics[width=\textwidth]{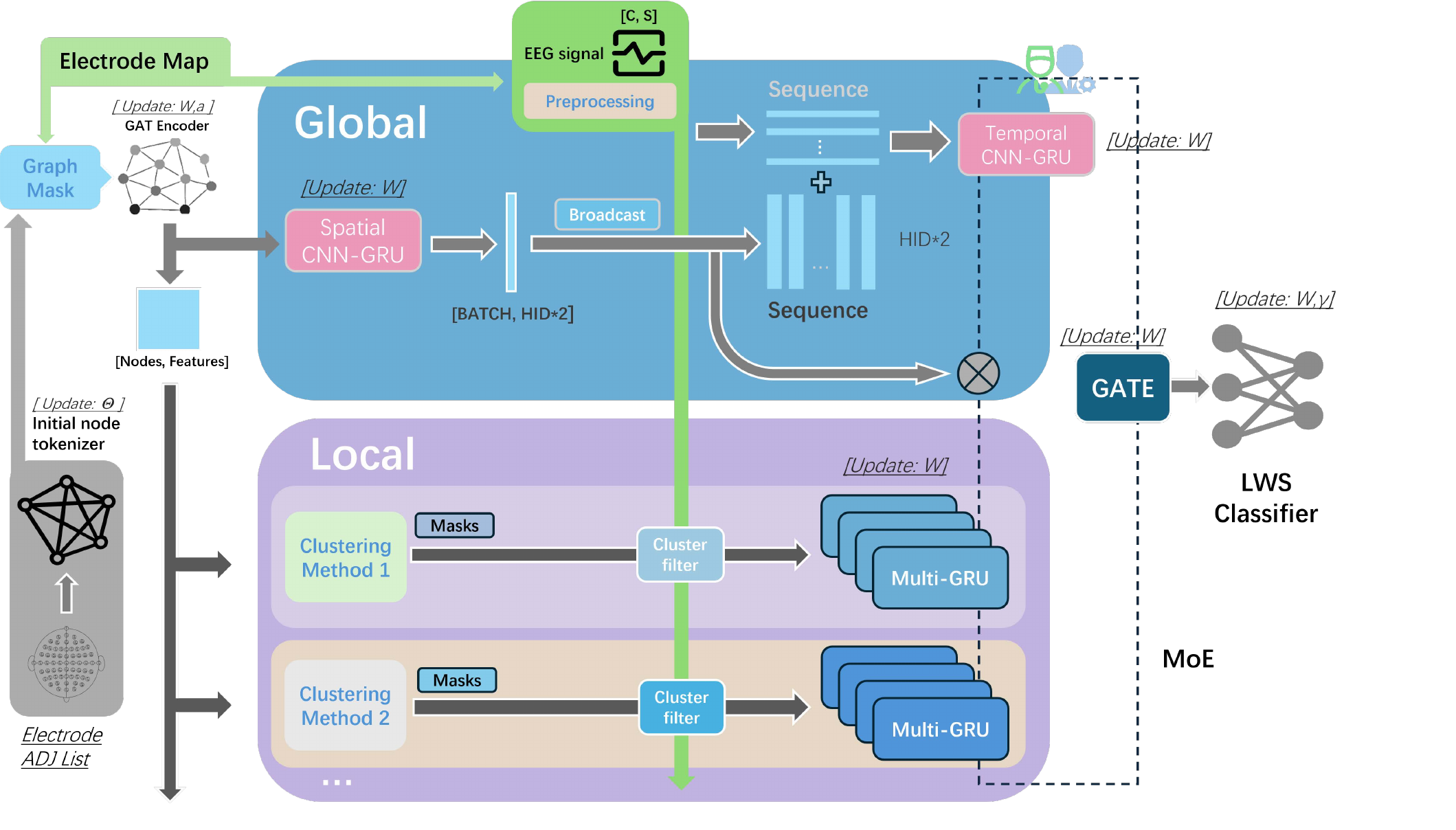} 
    \caption{Overview of the proposed \textbf{GCMCG} framework. The green modules denote real input signals and electrode metadata. Learnable parameters are annotated in light gray with \textit{Update:~}$\cdot$ for each module.}
	\label{fig:03_01_GCMCGframework}
\end{figure*}

\section{Method}\label{sec:Method}

We propose \textbf{GCMCG}, 
a graph-guided, clustering-aware mixture-of-experts CNN-GRU framework designed 
to jointly model the spatial dependencies 
and temporal dynamics inherent in EEG signals for robust decoding. 
As illustrated in~\autoref{fig:03_01_GCMCGframework}, 
the model takes two parallel inputs: the raw EEG signals 
and a spatial token representation derived from electrode topology. 
The EEG signals first undergo our ICA-WT preprocessing~(\ref{sec:3_1})
to enhance quality and suppress physiological and environmental artifacts, 
and are then forwarded to a global CNN-GRU backbone for initial temporal modeling, which captures holistic spatiotemporal patterns. 
Meanwhile, spatial tokens are initialized by a learnable tokenization module that encodes electrode positions into a structured graph, with a graph mask introduced to accommodate different electrode maps; these tokens are then processed by a graph-attention encoder to learn adaptive inter-electrode connectivity~(\ref{sec:3_2}).
The resulting graph is segmented into functional regions via an unsupervised clustering module. 
While our framework supports various clustering algorithms, 
we adopt spectral clustering in this study for its ability to infer the optimal number of regions automatically.
To enabling flexible and trainable region-to-expert assignment,
the discrete clustering results are paired with a learnable soft mask.
The clustered topology guides how the preprocessed EEG signals are routed to localized GRU experts for region-specific temporal modeling~(\ref{sec:3_3}). This procedure is crucial for dynamically establishing the link between EEG signal fluctuations from various brain regions and human movements.
Then, to integrate the outputs from the global and local networks, we employ a Mixture-of-Experts (MoE) system designed to capture the spatio-temporal patterns of EEG signals. This architecture allows local networks to specialize in detecting movement-specific scalp potentials from distinct brain regions, while the global network maintains a comprehensive integrative context. Within this architecture, we implement a stable, generalizable training strategy—combining focal loss, progressive sampling, and learnable scaling—to address class imbalance and inter-subject variability, thereby enhancing cross-subject robustness~(\ref{sec:3_4}).

%%To address limitations in prior GNN-based EEG models—including static graph assumptions, 
%%rigid inputs, and limited generalizability—GCMCG employs a learnable GAT encoder 
%%that adaptively captures inter-electrode dependencies without requiring fixed topologies. 
%%To enhance spatial interpretability and specialization, 
%%spectral clustering is applied to the learned embeddings to identify functional regions, 
%%each handled by a CNN-GRU expert. A gated fusion module then dynamically weights expert outputs, 
%%improving robustness to varying electrode layouts and temporal dynamics. 
%%As previously mentioned, the limitations of prior GNN-based EEG decoders include reliance on static graph structures, limited input flexibility, and insufficient generalization. Our GCMCG framework addresses these challenges through a unified approach that integrates adaptive graph learning with spectral clustering to dynamically identify functional brain regions. These regions are processed by specialized GRU experts, whose outputs are intelligently fused with a global temporal model via a gated MoE mechanism. Coupled with a robust training strategy, this design enables powerful and generalizable spatio-temporal decoding of EEG signals under variable conditions.

The learnable parameters of GCMCG are shown in~\autoref{tab:03_01_params}.
Throughout this paper, we follow the following notation conventions: 
boldface letters (e.g., $\mathbf{X}$, $\mathbf{W}$) denote vectors or matrices; 
calligraphic letters (e.g., $\mathcal{J}$, $\mathcal{G}$) indicate mappings, sets, or structured constructs; 
lowercase normal letters (e.g., $t$, $x$) are used for scalar variables, time steps, or function values; 
uppercase normal letters (e.g., $C$, $F$) represent constants, total counts or dimensions.

\begin{figure}[htbp]
	\centering
	\includegraphics[width=0.4\columnwidth]{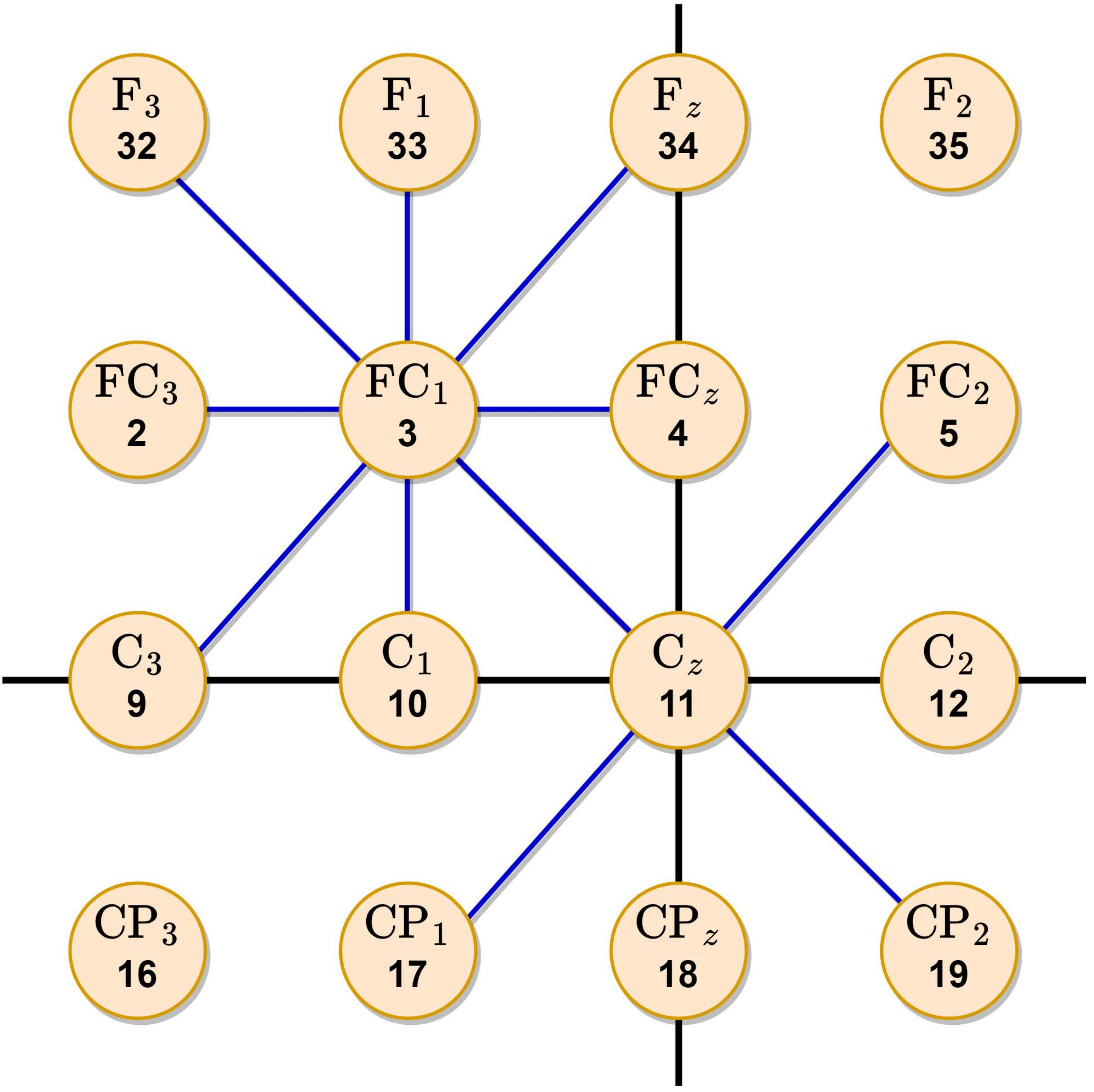} 
    \caption{Eight-connected graph structure centered on $\text{FC}_1$ and $\text{C}_z$, illustrating spatial electrode neighborhoods used for EEG graph construction.}
	\label{fig:03_02_eight}
\end{figure}

\subsection{Preprocessing}\label{sec:3_1}

Given raw EEG signals $\mathbf{X} \in \mathbb{R}^{C \times S}$, 
where $C$ is the number of channels and $S$ is the number of time steps, 
we process two types of inputs: 
the EEG signal itself and graph-based spatial information derived from a tokenization module. 
The signal input undergoes preprocessing that combines frequency-domain filtering 
with ICA-WT denoising to remove physiological and environmental artifacts. 
And the graph-based input initializes spatial information with a learnable, topology-aware tokenizer, producing adaptive and parameterized input representations beyond fixed graphs; a graph mask is further introduced to accommodate heterogeneous electrode maps.

%\subsubsection{EEG Signal Artifact Removal}\label{sec:3_1_1}
\begin{figure*}[htbp]
	\centering
	\includegraphics[width=\textwidth]{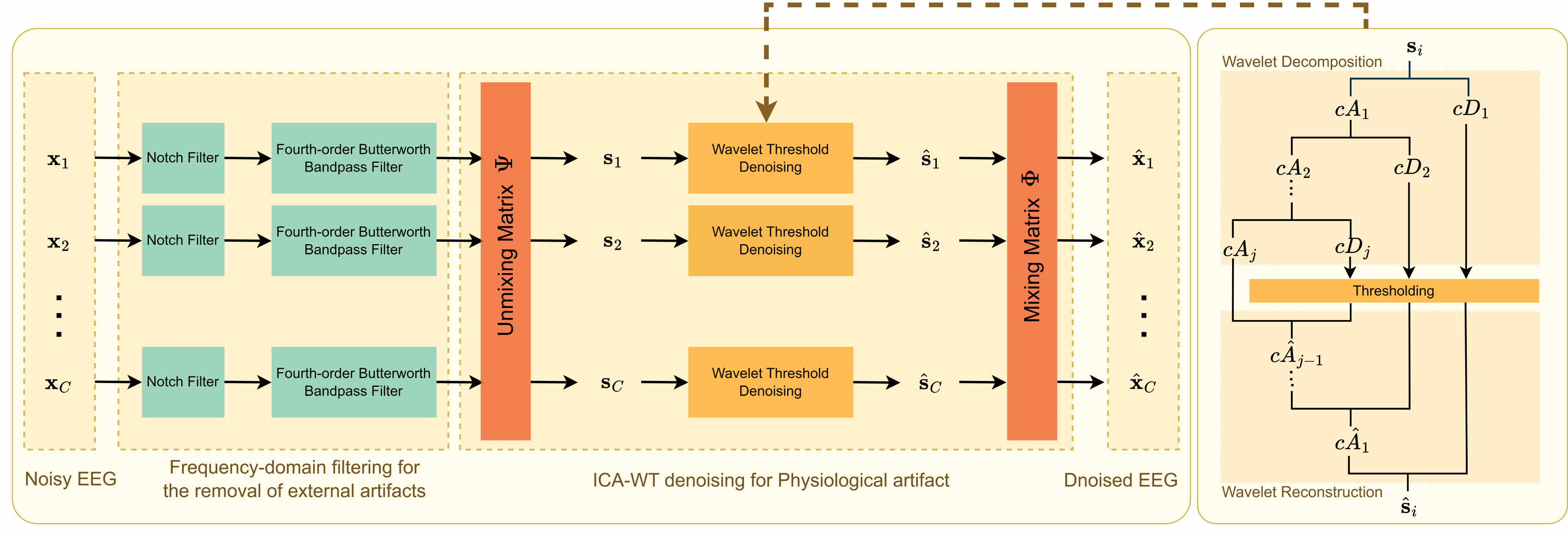} % 替换为你的图片文件名
	\caption{Schematic Diagram of the Hybrid Denoising Method. In the first stage, a notch filter and a second-order Butterworth bandpass filter are applied to eliminate external artifacts. In the second stage, physiological artifacts are eliminated using the ICA-WT algorithm (where $cA$ denotes wavelet approximation coefficients and $cD$ represents wavelet detail coefficients).}
	\label{fig:04_1_denoise}
\end{figure*}
% The hybrid denoising approach employing multiple filters for EEG signal is illustrated in~\autoref{fig:04_1_denoise}. 
%sec1:引入EEG信号去噪重要性+EEG信号去噪的挑战+引出所采用的方法
%sec2:介绍频域去噪

The hybrid denoising framework for EEG signals is illustrated in Fig.~\autoref{fig:04_1_denoise}. The process consists of two main stages. In the first stage, consistent with conventional EEG preprocessing, a notch filter and a second-order Butterworth bandpass filter are applied to eliminate external artifacts arising from environmental, instrumental, and measurement sources. However, due to spectral aliasing, frequency-domain filtering alone is insufficient to remove physiological artifacts from EEG signals \cite{mannan2018identification}. To address this limitation, the preprocessed signal is subsequently fed into the second stage, which employs the ICA-WT denoising algorithm inspired by \cite{al2017automaticICA}. This step constitutes a key contribution of our preprocessing pipeline. Through a structured procedure involving signal decomposition, wavelet-based denoising, and signal reconstruction, the method effectively suppresses noise across multiple independent components. This capability enables reliable physiological artifact removal in complex biological scenarios and produces a cleaned EEG signal with maximally preserved neural information and a significantly enhanced signal-to-noise ratio.

% This hybrid strategy combines the separation power of ICA with the denoising precision of WT, 
% making it well-suited for physiological artifact removal in EEG signals under complex biological conditions. 
The application of ICA requires that the target signals meet two conditions: 
1.~the independent components (ICs) are statistically independent and exhibit non-Gaussian distributions, 
and 2.~the target signals are linear mixtures of these independent components. 
It has been verified that EEG signals satisfy these two conditions\cite{james2004independent}. 
In the following, 
we will detail the denoising procedure, 
beginning with ICA. 
For a multi-channel EEG signal represented by the 
matrix $\mathbf{X}=[\mathbf{x}_1, \mathbf{x}_2, \ldots, \mathbf{x}_C]^\top$, 
wherein each row $\mathbf{x}_i^\top \in \mathbb{R}^{S}$ corresponds to 
a specific single-channel signal with length $S$. With the implementation of ICA, EEG signals $\mathbf{X}$ could be decompose into independent components $\mathbf{S}=[\mathbf{s}_1, \mathbf{s}_2, \ldots , \mathbf{s}_C]^\top$ with the ICA condition 2,  
%The ICA establishes a mathematical formulation describing the mixing and component processes as follows:

\begin{equation}\label{icad}
	\mathbf{X} = \mathbf{\Phi}\mathbf{S},
\end{equation}
in which the mixing matrix $\mathbf{\Phi} \in \mathbb{R}^{C \times C}$ could be estimated efficiently through Fast ICA algorithm\cite{hyvarinen2000independent}.

\par After the ICA decomposition algorithm is completed, 
we perform denoising on each component of the independent components $\mathbf{S}$. 
This process is divided into two steps: 
1.~Calculate the kurtosis (Kurt) of each independent component, 
as signals with lower kurtosis are closer to a Gaussian distribution. We set a threshold of 0.5 
and zero out the independent components with a kurtosis below that threshold. 
2.~Apply wavelet shrinkage denoising method\cite{donoho1994ideal} to the remaining independent components. 
The wavelet basis function chosen is db4, and the decomposition level is $j=\log_2(length(\mathbf{s}_i))$. 
%This denoising algorithm employs a fixed threshold to perform soft-thresholding on the wavelet coefficients of each layer obtained after wavelet decomposition of each independent component. 
This denoising algorithm employs a fixed threshold to shrink the wavelet coefficients $c_D$ of each layer obtained after wavelet decomposition of each independent component. Specifically, we use the following thresholding function:
\begin{equation}
\hat{c}_D =
\begin{cases}
c_D \left( 1 - e^{\frac{\lambda_T - c_D}{1.5}} \right), & c_D > \lambda_T, \\[6pt]
0, & |c_D| \leq \lambda_T, \\[6pt]
(-c_D)\left( e^{\frac{\lambda_T + c_D}{1.5}} - 1 \right), & c_D < -\lambda_T,
\end{cases}
\end{equation}
where $\lambda_T$ denotes the threshold.
This time-frequency domain denoising approach can effectively avoid impacting the effective components of the signal during the denoising process.
For each denoised independent component,
the inverse discrete wavelet transform (IDWT) is applied, 
resulting in the denoised independent components $\hat{\mathbf{S}}$. 
Subsequently, the following reconstruction algorithm 
is executed to obtain the denoised EEG signal $\hat{\mathbf{X}}$,

\begin{equation}\label{ICAD}
	\hat{\mathbf{X}} = \mathbf{\Phi} \hat{\mathbf{S}}.
\end{equation}

%sec4:介绍Normalization
%Finally, to ensure a stable learning rate for the model and to avoid overfitting issues, 
%each channel of the EEG signal is individually subjected to Standardization 
Finally, each channel of the EEG signal is individually subjected to standardization 
using mean removal and variance scaling, as shown in the following formula:
\begin{equation}
	\mathbf{z}_i = \frac{\hat{\mathbf{x}}_i - \mu}{\sigma}, \ \quad i \in \{1, 2, \dots, C\},
\end{equation}
where,
$\mathbf{z}_i$ represents the standardized single-channel EEG signal,
$\hat{\mathbf{x}}_i$ denotes the denoised single-channel EEG signal,
$\mu$ is the mean of $\hat{\mathbf{x}}_i$, 
and $\sigma$ signifies the standard deviation of $\hat{\mathbf{x}}_i$.
And the $\mathbf{Z} =[\mathbf{z}_1, \mathbf{z}_2, \ldots, \mathbf{z}_C]^\top$ is the standardized EEG signal matrix.

%\subsubsection{Spatial Encoding}\label{sec:3_1_2}
With the cleaned and standardized signal prepared, a companion spatial branch encodes the electrode topology. We build an adjacency-list set \(\mathcal{J}\) in which each node \(i\) is assigned neighbors \(\mathcal{J}(i)\) following an eight-connected layout~(\autoref{fig:03_02_eight}).
Leveraging \(\mathcal{J}\), we initialize the learnable node features \(\mathbf{\Theta}^0 \in \mathbb{R}^{C \times F}\) with an electrode-aware tokenizer that encodes the scalp layout (where \(F\) denotes the feature length). This initialization injects spatial priors that steer the GAT encoder even under random weights; an optional graph mask further accommodates heterogeneous electrode montages.

Taken together, the preprocessing stage yields two synchronized inputs: (i) the denoised, standardized EEG signal, which is fed to a global CNN–GRU backbone for holistic temporal feature extraction; and (ii) the graph-structured spatial tokens, which are independently encoded by a GAT. The two information paths are fused downstream via expert assignment and gating, where clustered graph representations act as routing signals for localized temporal modeling.

\subsection{Graph Attention Encoding and Electrode Clustering}\label{sec:3_2}

We employ GAT~\cite{velivckovic2017GAT}, 
a message-passing architecture where each node aggregates information from its neighbors using learned attention weights. 
While $\mathcal{G}_\text{masked} \in \mathbb{R}^{C \times F}$ is the initial node feature matrix (masked tokenizer set),
for a given node $i$, its output feature is computed as:

\begin{equation}
\mathbf{\Theta}_i^{t} = f_\text{GAT}^t(\sum_{j \in \mathcal{J}(i)} \alpha_{ij}^t \mathbf{W_{\text{GAT}}} \mathbf{\Theta}_j^{t-1})
\end{equation}

Here, $t$ indicates the layer,
$f$ is the activation function,
$\mathbf{\Theta}_j$ is the feature of neighboring node $j$, 
$\mathbf{W_{\text{GAT}}}$ is a learnable weight matrix, 
and $\alpha_{ij}$ is the normalized attention coefficient between node $i$ and $j$ computed as:

\begin{equation}
\alpha_{ij}^t = \frac{\exp \left( \text{LeakyReLU} \left( \mathbf{a}^{\top} [\mathbf{W_{\text{GAT}}} \mathbf{\Theta}_i \,\|\, \mathbf{W_{\text{GAT}}} \mathbf{\Theta}_j] \right) \right)}{\sum_{k \in \mathcal{J}(i)} \exp \left( \text{LeakyReLU} \left( \mathbf{a}^{\top} [\mathbf{W_{\text{GAT}}} \mathbf{\Theta}_i \,\|\, \mathbf{W_{\text{GAT}}} \mathbf{\Theta}_k] \right) \right)},
\end{equation}
where $\mathbf{a}$ is a learnable attention vector and $\|$ denotes vector concatenation. Multi-head attention is used to stabilize learning and improve expressiveness.
The final graph-enhanced embedding is $\mathbf{\Theta'} \in \mathbb{R}^{C \times D}$, where $D$ is the embedding dimension in hidden layers.
The resulting $\mathbf{\Theta'}$ is used for downstream electrode clustering and expert assignment.

%\subsection{Electrode Clustering}

To leverage regional patterns in brain activity, 
We adopt spectral clustering to adaptively infer the number of clusters using the eigengap heuristic, 
presented in Algorithm~\ref{alg:03_02_clustering}. 
We first compute the correlation matrix from GAT-learned node embeddings, 
then apply Laplacian eigen decomposition and select the optimal $K$ via the eigengap criterion. 
This process yields a discrete assignment of each EEG channel to a cluster index, 
which guides the expert dispatching module in downstream processing.

This clustering design not only addresses the challenge of lacking biologically meaningful spatial priors, 
but also introduces a new modeling aspect—how to ensure architectural consistency under flexible region partitioning.
By assigning electrodes to functional subregions and dispatching them to corresponding GRU experts, 
the model can localize temporal modeling within spatially coherent areas. 
This promotes specialization across experts while preserving global coordination through gated fusion, 
enabling more efficient and interpretable integration of spatial topology and temporal dynamics.

\subsection{Multi-Branch GRU Expert Module}\label{sec:3_3}

Based on the clustered subregions and their corresponding channel assignments, the preprocessed EEG signals from each region are routed to a dedicated GRU expert.
The GRU combines the forget and input gates into a single update gate, 
and merges the cell state and hidden state. 
The core computations of GRU are described in~\ref{equ1:gru}.

While local GRU experts specialize in region-specific temporal patterns, 
a separate global CNN-GRU branch captures cross-regional dependencies, 
thereby complementing the localized pathways. 
To jointly capture spatial and temporal dependencies, 
the model employs three types of GRU experts: 
a spatial GRU over the GAT output, 
a temporal GRU over the raw EEG sequence concatenated with graph-informed embeddings, 
and $K$ cluster-specific GRU experts. 
Each cluster expert receives masked input data corresponding to its assigned electrodes.
This multi-branch architecture enables the model to extract diverse spatial-temporal features at varying resolutions: 
the spatial GRU learns diffusion-like transitions across electrode neighborhoods encoded by GAT, 
the temporal GRU emphasizes global sequential dynamics enhanced by topological context, 
while the cluster-specific experts perform fine-grained temporal modeling within localized functional regions. 
By decoupling these learning pathways, 
GCMCG improves representational capacity, 
and enhances interpretability by associating specific latent patterns with spatial priors.

The GRU implementation combines a 1D convolutional front-end for initial feature extraction with a gated recurrent backbone:
\begin{equation}
    \mathbf{v}_k = \text{GRU}_k(\text{Conv1D}(\mathbf{Z}_k)) \in \mathbb{R}^{2D}, \quad k \in \{1, 2, \dots, K\}
\end{equation}
where $k$ indexes the cluster or expert branch, 
$\mathbf{Z}_k$ is the masked input data for the $k$-th expert,
and $2D$ is the GRU output dimension.
%\subsection{Adaptive Expert Fusion with Entropy-Regularized Gating}
The outputs from all expert branches are aggregated via a learnable gating mechanism. 
Let the concatenated expert features (fusion) be:
\[
\hat{\mathbf{v}} = \mathbf{v}_{\text{spatial}} \,\|\, \mathbf{v}_{\text{temporal}} \,\|\, \mathbf{v}_1 \,\|\, \ldots \,\|\, \mathbf{v}_K, \quad \hat{\mathbf{v}} \in \mathbb{R}^{(K+2) \cdot 2D}
\]
we first compute unnormalized gating logits via a shared linear projection:
\begin{equation}
    \mathbf{g}^t = \exp(f^t_\text{gate}(\mathbf{w}^t_\text{gate}{\mathbf{g}^{t-1}} + b^t_\text{gate})), \ \mathbf{g}^0 = \hat{\mathbf{v}}
\end{equation}
where $t$ is the layer, $\mathbf{w}_{gate}$ and $b_{gate}$ are learnable parameters. 
The output gating logits is $\mathbf{g}', \ \mathbf{g}' \in \mathbb{R}^{K+2}$.
The normalized gating weights (gate score) are then computed as:
\begin{equation}
    \boldsymbol{\alpha}_\text{gate} = \frac{\mathbf{g}'}{1 + \sum_{j=1}^{K+2} \mathbf{g}'_j}
\end{equation}
An entropy-based regularization term is added to promote sparsity and expert specialization:
\begin{equation}
    \mathcal{L}_{\text{gate}} = -\frac{1}{B} \sum_{b=1}^{B} \sum_{i=1}^{K+2} \boldsymbol{\alpha}_{b,i} \log(\boldsymbol{\alpha}_{b,i} + \epsilon)
\end{equation}
where $B$ is the batch size and $\epsilon$ is a small constant for numerical stability.
The final fused representation is then computed as a weighted sum of expert outputs:
\begin{equation}
    \hat{\mathbf{v}}_{\text{fused}} = \sum_{i=1}^{K+2} \boldsymbol{\alpha}_i \cdot \hat{\mathbf{v}}_i
\end{equation}

The final representation is obtained via weighted sum fusion, 
and an entropy-based regularization term is applied to encourage sparsity 
and specialization among experts. 
%The detailed procedure is outlined in~\ref{alg:03_03_fusion}.

\subsection{Classification Head and Training Strategy}\label{sec:3_4}
\begin{algorithm}[htbp]
    \caption{Classification and Three-Stage Training Strategy}\label{alg:03_04_lws_training}
    \begin{algorithmic}[1]
    \REQUIRE{Fused features $\hat{\mathbf{v}}_{\text{fused}} \in \mathbb{R}^{2D}$, class labels $\mathbf{y}$, training epochs $E$}
    \ENSURE{Final classification logits and trained LWS head}
    
    \STATE{\textbf{Phase 1: Pretraining with Cross Entropy loss}}
    \STATE \quad Train full LWS head with standard CE loss and random sampler
    
    \STATE \textbf{Phase 2: Fine-tuning with focal loss and PBS}
    \STATE \quad Freeze backbone parameters
    \STATE \quad Replace sampler with \textbf{ProgressivelyBalancedSampler}
    \STATE \quad Replace loss with \textbf{FocalLoss}
    
    \STATE \textbf{Phase 3: Final scaling}
    \STATE \quad Freeze all FC layers in classifier
    \STATE \quad Enable LWS: train only the learnable scaling vector $\boldsymbol{\gamma}$
    \STATE \quad Use CE or focal loss for refinement
    
    \RETURN Scaled logits: $\hat{\mathbf{y}} = \mathbf{\hat{W}} \cdot \boldsymbol{\gamma} \cdot \mathbf{Z}_{\text{fused}} + \mathbf{\hat{b}}$
    \end{algorithmic}
\end{algorithm}

The fused representation is passed to 
a learnable weight scaling~(LWS) classification 
head~\cite{kang2019decoupling}, 
which consists of several fully connected layers and a final output layer. 
The LWS mechanism applies per-class scaling to handle data imbalance and 
improve interpretability. To further address class imbalance, 
we adopt a three-stage training strategy: 
(1) pretraining with Cross Entropy~(CE) loss, 
(2) fine-tuning with focal loss~\cite{lin2017focal} 
and Progressively Balanced Sampling (PBS), and 
(3) final adjustment of LWS parameters with frozen backbone.
The overall classification and training strategy is summarized in Algorithm~\ref{alg:03_04_lws_training}.
Based on $\hat{\mathbf{v}}_{\text{fused}}$, we can compute the final logits as $\hat{y}$.
To jointly optimize classification performance and expert sparsity, we define the classification loss as:
\[
\ell_{\mathrm{cls}}(\mathbf{y},\mathbf{p})
= -\sum_{q=1}^Q \mathbf{y}^{(q)}\log \mathbf{p}^{(q)},
\quad
\mathbf{p}_{n}=\mathrm{Softmax}(\hat{\mathbf{y}}_{n}),
\]
where $Q$ is the total number of classes,
then the overall loss is:
%\begin{equation}\label{eq:joint-loss}
%    \mathcal{L}
%    =
%    \underbrace{\frac{1}{N}\sum_{n=1}^N \ell_{\mathrm{cls}}\bigl(\mathbf{y}^{(n)},\,\mathbf{p}^{(n)}\bigr)}_{\text{Classification Loss}}
%    \;+\;
%    \underbrace{
%        \lambda_{\text{gate}} \cdot \mathcal{L}_{\text{gate}}
%      }_{\text{Entropy Regularization}} 
%\end{equation}
\begin{equation}
    \mathcal{L} = \frac{1}{N} \sum_{n=1}^{N} \ell_{\mathrm{cls}}\left(\mathbf{y}_{n}, \mathbf{p}_{n}\right) 
    + \lambda_{\mathrm{gate}} \cdot \mathcal{L}_{\mathrm{gate}}
    \label{eq:overall-loss}
\end{equation}    
where $\lambda_\text{gate}$ is a hyper-parameter used for entropy regularization, 
$N$ is the total number of samples,
and $n$ is the index of the current sample. 
This overall loss encourages a low-entropy gating distribution. 

\subsection{Summary}
The proposed GCMCG model integrates spatial graph modeling, 
functional clustering, dynamic masking, 
and expert fusion into a unified and flexible decoding pipeline. 
It is designed to generalize across datasets, 
subject variations, and channel configurations, 
making it suitable for practical MI/ME BCI applications.
The previously introduced non-learnable symbols are summarized in~\autoref{tab:03_03_notations}.

\section{Experiments and results}\label{sec:Experiments}

To comprehensively evaluate the performance of GCMCG, we first conducted an extensive comparative analysis against a wide spectrum of deep learning baselines, including convolutional neural network variants (EEGNet~\cite{lawhern2018eegnet}, CMO-CNN~\cite{LIU2023CMOCNN}, FFCL~\cite{li2022FFCL}) and recurrent models (TD-LSTM~\cite{karimian2024tdlstm} and RNN-GRU~\cite{luo2018rnnGRU}). This broad comparison was designed to benchmark overall model performance, highlighting the limitations of existing methods in multi-classification and cross-subject generalization while demonstrating the advantages of GCMCG. Subsequently, we performed ablation studies to systematically validate the contributions of key components. We assessed the effectiveness of our robust denoising and adaptive input pipeline, which employs a hybrid denoising method to mitigate different noise, confirming its critical role in improving data quality. The graph attention module was also examined through ablation, verifying its importance for spatial specialization within our unified spatio-temporal framework. Furthermore, to explore the interpretability of our model's spatial-temporal structure, we visualized t-SNE embeddings and illustrated the graph connectivity of electrodes, revealing the synergistic feature learning achieved through the integration of graph attention networks with spectral clustering and gated multi-expert fusion. Finally, we evaluated the efficacy of the three-stage training strategy, which incorporates focal loss, progressive sampling, and learnable scaling, thereby validating its role in enhancing cross-subject robustness and overall training stability.

The experiments were conducted on an Ubuntu 24.04, 
64-bit machine with a 2.8 GHz AMD CPU and 
an Nvidia A100 GPU used 
to train and test the model. 
Python 3.12.3 and Pytorch 2.5.1+cu121
is used to implement 
the suggested method. 
All data are calculated using tensor structure, 
caused by the computational efficiency.

All experiments are conducted using three publicly available EEG datasets: 
EEGmmidb (BCI2000)~\cite{schalk2004bci2000}, 
BCI Competition IV-2a~\cite{brunner2008bci}, 
%BCI Competition %IV-2b~\cite{leeb2008bci}, 
and the M3CV~\cite{M3CV} dataset. 
The goal of utilizing these diverse datasets is to systematically evaluate 
the generalizability and robustness of the proposed method 
across multiple paradigms and recording conditions. 
Crucially, all selected datasets include both MI and ME tasks, 
enabling us to investigate the shared and distinct neural representations 
between imagined and actual movements, 
which is essential for practical BCI applications.
The inclusion of these datasets also reflects 
the need to capture variations in subject populations, 
session numbers, electrode configurations, and task complexities. 
While BCI-IV 2a focuses on 
four-class MI with high inter-subject variability, 
EEGmmidb extends this to nine classes involving 
both MI and ME in a more naturalistic BCI setting. 
The M3CV dataset provides multi-session recordings from over 100 subjects across various EEG tasks. However, for consistency, here only the MI and ME paradigms are used in M3CV.

All models are trained using leave-one-subject-out (LOSO) 
or leave-group-of-subjects-out (LGSO) cross-validation strategies to evaluate cross-subject generalization. 
Specifically, LOSO is applied to BCI Competition IV-2a datasets, 
where each subject is left out in turn for testing. 
For the EEGmmidb and M3CV datasets, which include a larger number of subjects, 
we adopt the LGSO protocol by partitioning subjects into multiple groups 
and leaving one group out for testing at each iteration.
The underlying principles of LOSO and LGSO are essentially equivalent: 
both aim to assess the model's ability to generalize to unseen subjects. 
As illustrated in Algorithm~\ref{alg:04_05_lgso},
when the number of groups in LGSO equals the number of subjects, LGSO becomes identical to standard LOSO.

\begin{algorithm}[htbp]
    \caption{Leave-Group-of-Subjects-Out (LGSO) Cross-Validation}
    \label{alg:04_05_lgso}
    \begin{algorithmic}[1]
    \REQUIRE Total subject set $\mathcal{S}$, total groups $N$, current group index $m$
    %\ENSURE Training and test splits: $(\mathcal{X}_{\text{train}}, \mathcal{y}_{\text{train}})$, $(\mathcal{X}_{\text{test}}, \mathcal{y}_{\text{test}})$
    
    \STATE Sort subject IDs in $\mathcal{S}$
    \STATE Compute group size $s = \lfloor |\mathcal{S}| / N \rfloor$
    \STATE Distribute remaining subjects to the first $(|\mathcal{S}| \bmod N)$ groups
    \STATE Partition $\mathcal{S}$ into $G$ disjoint subsets: $\{\mathcal{S}_1, \dots, \mathcal{S}_n\}$
    
    \STATE Define test subjects: $\mathcal{S}_{\text{test}} = \mathcal{S}_m$
    \STATE Define train subjects: $\mathcal{S}_{\text{train}} = \mathcal{S} \setminus \mathcal{S}_{\text{test}}$
    
    \STATE Concatenate training features and labels:
    \[
    \mathcal{X}_{\text{train}} = \bigcup_{s \in \mathcal{S}_{\text{train}}} \mathcal{X}^{(s)}, \quad
    \mathcal{y}_{\text{train}} = \bigcup_{s \in \mathcal{S}_{\text{train}}} \mathcal{y}^{(s)}
    \]
    
    \STATE Concatenate test features and labels:
    \[
    \mathcal{X}_{\text{test}} = \bigcup_{s \in \mathcal{S}_{\text{test}}} \mathcal{X}^{(s)}, \quad
    \mathcal{y}_{\text{test}} = \bigcup_{s \in \mathcal{S}_{\text{test}}} \mathcal{y}^{(s)}
    \]
    
    \RETURN $(\mathcal{X}_{\text{train}}, \mathcal{y}_{\text{train}})$, $(\mathcal{X}_{\text{test}}, \mathcal{y}_{\text{test}})$
    \end{algorithmic}
    \end{algorithm}

\subsection{Dataset and Preprocessing Output}
%==============================================
%在此处介绍所使用的数据集并引出下文处理
%==============================================
The PhysioNet EEG Motor Movement/Imagery dataset was acquired 
using the BCI2000 system and is accessible through 
the \href{https://physionet.org/content/eegmmidb/1.0.0/}{eegmmidb/1.0.0}\cite{schalk2004bci2000, goldberger2000physiobank}. 
This dataset comprises over 1500 EEG recordings, 
each lasting one or two minutes, collected from 109 subjects. 
Each subject participated in 14 experiments, 
which can be categorized into five types: baseline runs (eyes open and eyes closed), 
open and close left or right fist, 
imagine opening and closing left or right fist, 
open and close both fists or both feet, 
and imagine opening and closing both fists or both feet. 
Each experiment comprises a total of 29 trials, 
with each trial lasting approximately 4 seconds. 
The signal sampling rate is 160 Hz, and the number of signal channels is 64.
%\subsection{Dataset Preprocessing}
%sec1:介绍BCI2000的数据切割
For the EEG Motor Movement/Imagery dataset, 
excluding the first two one-minute baseline tasks (open and close eyes) for each subject, 
each signal needs to be segmented according to the 29 trials` categories, 
and each segment should be assigned to a specific category. 
In total, there can be up to 9 categories, including:
T0 (rest), Imagery Both Fists, Imagery Both Feet, Movement Both Fists, 
Movement Both Feet, Imagery Left Fist, Imagery Right Fist, Movement Left Fist and Movement Right Fist. 
Additionally, taking into account the inherent variability in reaction speeds among subjects 
and to ensure consistency in data dimensions, only the first 4 seconds of each trial are retained, which amounts to 640 sample points. 
\begin{figure}[htbp]
	\centering
	\includegraphics[width=0.8\textwidth]{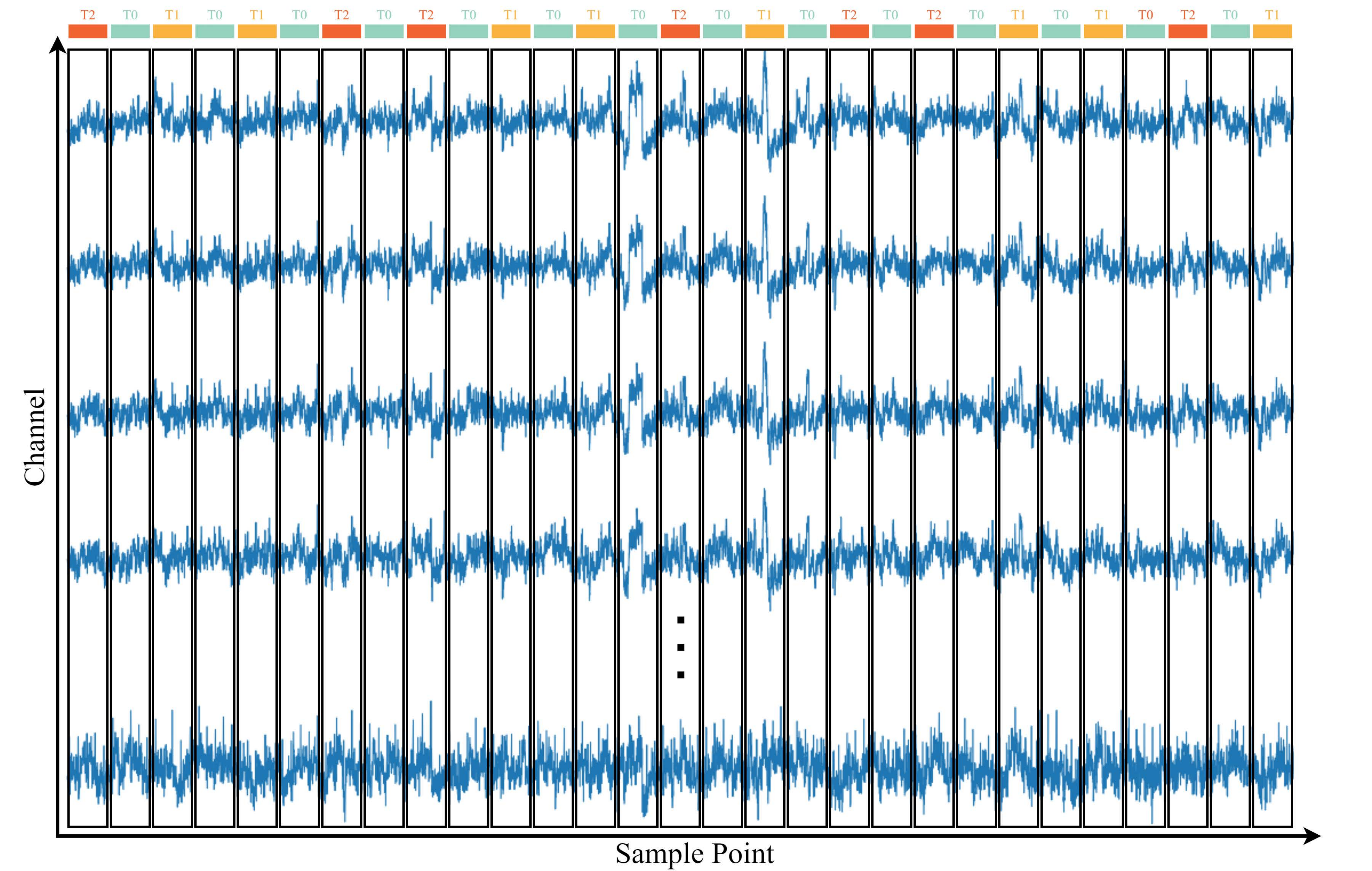} % 替换为你的图片文件名
	\caption{This figure illustrates the data-slicing process applied to the EEG recordings of subject S001 during Task 3 in the EEGmmidb dataset. The continuous EEG signals are sliced into segments and labeled according to the annotation file (T0: rest; T1: opening and closing of the left fist; T2: opening and closing of the right fist). Only the signals from the first four channels and the last channel are displayed.}
	\label{slice}
\end{figure}

A schematic diagram illustrating the use of a time window to segment and classify EEG signals is shown in \autoref{slice}.
In this project, for BCI2000 dataset, 
a 60Hz notch filter is applied to the EEG signals to eliminate power line interference. 
Subsequently, a fourth-order Butterworth bandpass filter is used to confine the signal within the range of 0.2Hz to 75Hz, 
thereby removing external artifacts originating from environmental, 
in strumental, and measurement artifacts.

\begin{figure}[htbp]
	\centering
	\includegraphics[width=0.8\textwidth]{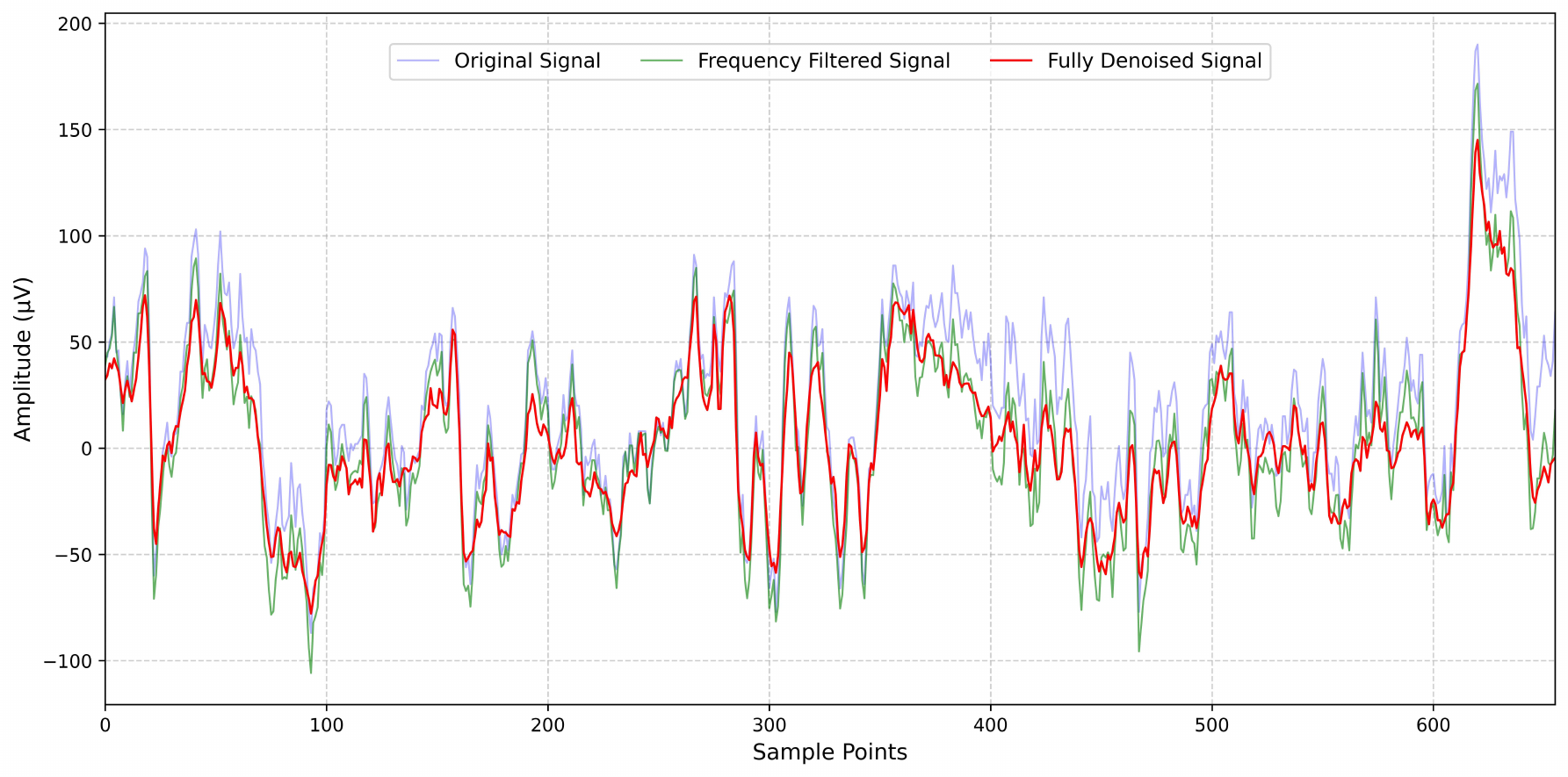} % 替换为你的图片文件名
	\caption{Illustration of EEG signal denoising using frequency-domain filtering followed by the ICA–WT hybrid method.
    A representative EEG channel is shown to compare the original signal (blue), the frequency-filtered signal (green), and the fully denoised signal (red). The sequential application of frequency filtering and ICA–WT denoising effectively suppresses noise while preserving the essential signal information.}
	\label{denoisePlt}
\end{figure}

The denoised single-channel EEG signal is shown in~\autoref{denoisePlt}.
As can be seen from the~\autoref{denoisePlt}, 
the denoised single-channel EEG signal (red line) 
compared to the original noisy signal (blue line, which has been amplitude-normalized for visualization), 
shows that the spikes in the signal have been significantly reduced, 
indicating an improvement in the signal-to-noise ratio (SNR) and effective removal of artifacts. 
Additionally, the baseline of the denoised signal consistently adheres 
closely to the original noisy signal without exhibiting excessive smoothing, 
suggesting that the effective information in the signal has been well preserved.

The BCI Competition IV Dataset 2a~\cite{brunner2008bci}, provided by Graz University of Technology, 
comprises EEG recordings from 9 subjects performing four-class MI tasks: left hand, right hand, both feet, and tongue. 
Each subject participated in two sessions conducted on different days. 
Each session consists of six runs, with each run containing 48 trials, totaling 288 trials per session. 
During each trial, subjects were visually cued to perform the specified MI task.
EEG signals were recorded using 22 Ag/AgCl electrodes positioned according to the international 10--20 system, 
covering motor and central cortical areas. 
Additionally, three monopolar electrooculogram (EOG) channels were recorded to monitor eye movements. 
The data were sampled at 250 Hz and bandpass filtered between 0.5 and 100 Hz, with a 50 Hz notch filter applied to suppress power line noise. 
This dataset is widely used for benchmarking motor imagery classification algorithms due to its standardized protocol and comprehensive recordings.
The recordings had a dynamic range of ±100 µV for the screening sessions and ±50 µV for the feedback sessions. 
This dataset is widely used for benchmarking binary motor imagery classification algorithms and studying session-to-session variability in BCI systems.

The M3CV (Multi-subject, Multi-session, Multi-task Commonality and Variability) 
dataset is a large-scale EEG database that covers 106 healthy participants with two recording sessions on different days~\cite{M3CV}. 
It includes six paradigms: resting-state, transient-state sensory (VEP, AEP, SEP), steady-state sensory (SSVEP, SSAEP, SSSEP), 
cognitive oddball (P300), motor execution (left hand, right hand, right foot), 
and SSVEP with selective attention. The recordings were made using 64-channel EEG with a sampling rate of 1000 Hz. 
This dataset provides diverse EEG tasks and supports cross-session, cross-task, and cross-subject evaluations, 
which is crucial for real-world BCI applications that require robust generalization. 
The motor execution part is been used in this study.

\begin{figure*}[htbp]
    \centering
    \hspace*{-3cm}  % 根据需要调整数值，向左移动图像
    \includegraphics[width=\textwidth]{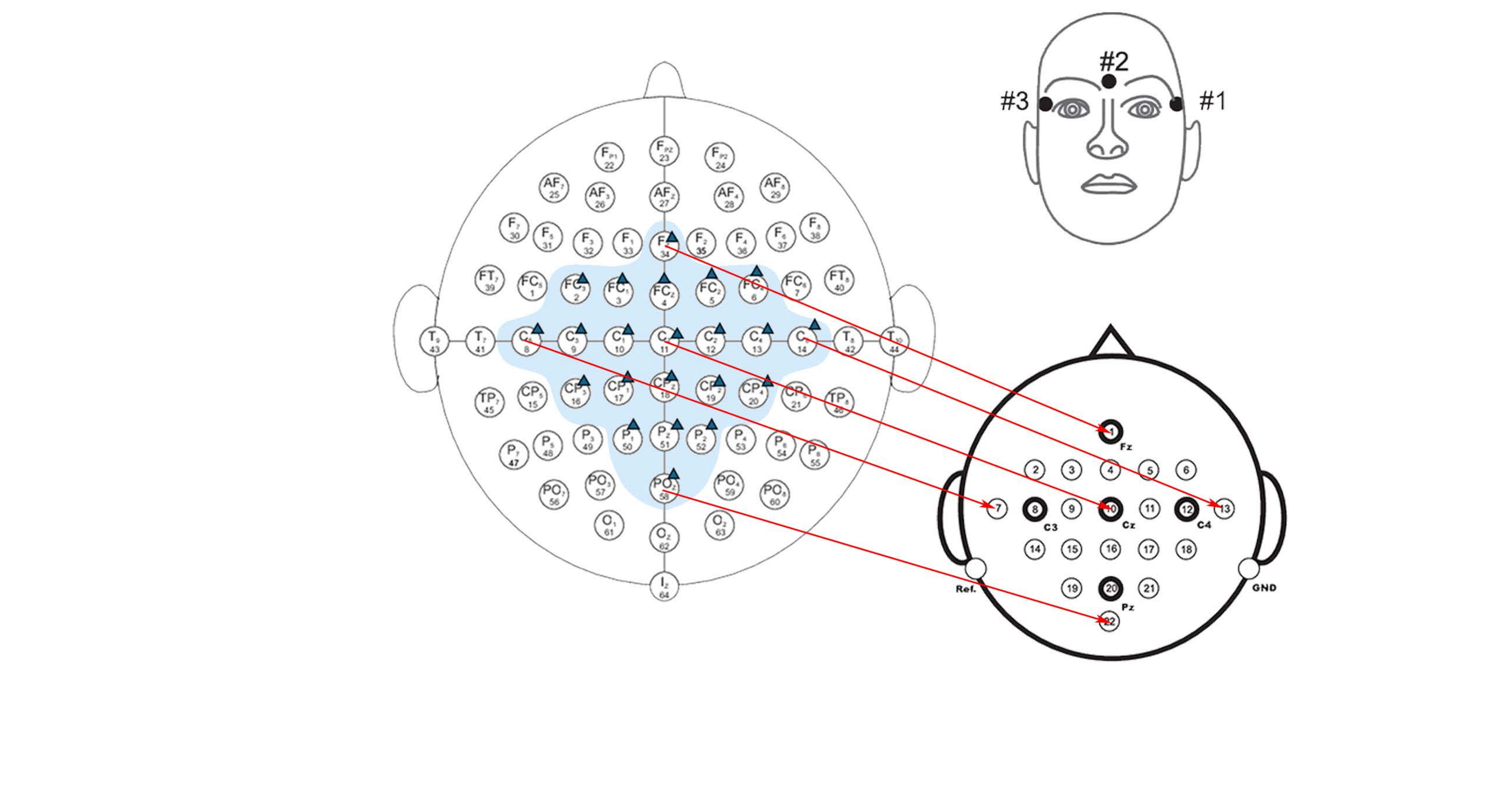} 
    \caption{Electrode layouts and graph mask construction. Top: 22-channel montage from BCI-IV dataset~\cite{brunner2008bci}. Bottom: 64-channel layout from BCI2000~\cite{schalk2004bci2000}, with selected nodes highlighted by the graph mask.}
    \label{fig:03_03_gmask}
\end{figure*}

Besides, based on the adjacency list, 
we construct the graph $\mathcal{G} = (\mathbf{\Theta}, \mathcal{J})$,
where $\mathbf{\Theta}$ is the node features.
We also provide a graph mask, constructed from a predefined electrode map (see \autoref{tab:03_02_adjMAP}b), to control which nodes participate more in graph modeling (see \autoref{fig:03_03_gmask}). This mechanism supports datasets with different channel layouts, provided their electrode designs reflect comparable functional topology.

\begin{table}[htbp]
    \captionsetup{labelfont=bf, textfont=normalfont, justification=raggedright, singlelinecheck=false}
    \caption{Graph input configuration excerpts}\label{tab:03_02_adjMAP}
    \subfloat[Adjacency list $\mathcal{J}$]{
        \begin{tabular*}{0.45\linewidth}{@{\extracolsep{\fill}} c l }
            \toprule
            Node & Neighbors \\
            \midrule
            1   & [2, 8, 9, 30, \\
                & 31, 32, 39, 41] \\
            2   & [1, 3, 8, 9, \\
                & 10, 31, 32, 3] \\
            $\vdots$ & $\vdots$ \\
            67  & [65, 66] \\
            \bottomrule
        \end{tabular*}
    }
    \quad
    \subfloat[Electrode mapping $\mathcal{E}$]{
        \begin{tabular*}{0.4\linewidth}{@{\extracolsep{\fill}} c l }
            \toprule
            Node & Mapping \\
            \midrule
            1   & 34 \\
            2   & 2 \\
            3   & 3 \\
            $\vdots$ & $\vdots$ \\
            24  & 66 \\
            25  & 67 \\
            \bottomrule
        \end{tabular*}
    }
\end{table}

  \begin{table}[htbp]
    \centering
    \caption{Comparison of the overall accuracy (Top1-Acc), macro-recall (Macro-Rec), macro-precision (Macro-Prec), Macro-F1 score and Cohen's Kappa (Kappa) across different models on EEGmmidb dataset}\label{tab:overall}
    \begin{tabularx}{\textwidth}{lXXXXXX}
      \toprule
      \textbf{Model} & \textbf{Top1-Acc (\%)} & \textbf{Macro-Rec (\%)} & \textbf{Macro-Prec (\%)} & \textbf{Macro-F1} & \textbf{Kappa}\\
      \midrule
      \textbf{GCMCG} & \textbf{86.60} & \textbf{79.41} & \textbf{83.07} & \textbf{0.81} & \textbf{0.82}\\
      FFCL~\cite{li2022FFCL} & 66.10 & 50.93 & 58.13 & 0.54 & 0.53\\
      CMO-CNN~\cite{LIU2023CMOCNN} & 60.38 & 43.75 & 46.88 & 0.45 & 0.47\\   %%
      EEGNet~\cite{lawhern2018eegnet} & 57.59 &  37.68 & 44.34 & 0.39 & 0.42\\
      GRU-RNN~\cite{luo2018rnnGRU} & 54.17 & 40.45 & 41.08 & 0.41 & 0.40\\
      TD-LSTM~\cite{karimian2024tdlstm} & 53.78 & 27.81 & 34.81 & 0.28 & 0.31\\
      \bottomrule
    \end{tabularx}
\end{table}

\subsection{Evaluation Criteria and Comparative results}

The overall accuracy (Top1-Acc) is defined as:
\begin{equation}
    \text{Top1-Acc} = \frac{\sum_{i=1}^{N} (TP_i + TN_i)}{\sum_{i=1}^{N} (TP_i + TN_i + FP_i + FN_i)}
\end{equation}
where \( TP_i \), \( TN_i \), \( FP_i \), and \( FN_i \) denote the numbers of true positives, 
true negatives, false positives, and false negatives for class \( i \), respectively.

The precision and recall for each class \( q \) are defined as:
\begin{equation}
    \text{Precision}^{(q)} = \frac{TP^{(q)}}{TP^{(q)} + FP^{(q)}}, \quad
    \text{Recall}^{(q)} = \frac{TP^{(q)}}{TP^{(q)} + FN^{(q)}}
\end{equation}

The macro-averaged precision and recall are then obtained by averaging over all classes:
\begin{equation}
    \text{Macro-Precision} = \frac{1}{Q} \sum_{q=1}^{Q} \text{Precision}^{(q)}, \quad
    \text{Macro-Recall}    = \frac{1}{Q} \sum_{q=1}^{Q} \text{Recall}^{(q)}
\end{equation}

The corresponding class-wise F1-score is:
\begin{equation}
    \text{F1}^{(q)} = 2 \times \frac{\text{Precision}^{(q)} \cdot \text{Recall}^{(q)}}{\text{Precision}^{(q)} + \text{Recall}^{(q)}}
\end{equation}

The macro F1-score is computed by averaging over all classes:
\begin{equation}
    \text{Macro-F1} = \frac{1}{Q} \sum_{q=1}^{Q} \text{F1}^{(q)}
\end{equation}

To further assess agreement beyond chance, Cohen’s kappa is computed as:
\begin{equation}
    \kappa = \frac{p_o - p_e}{1 - p_e}
\end{equation}
where \( p_o \) is the observed accuracy, and \( p_e \) is the expected agreement by chance, estimated by:
\begin{equation}
    p_e = \sum_{i=1}^{N} \left( \frac{(TP_i + FP_i) \cdot (TP_i + FN_i)}{N} \right)
\end{equation}

Table~\ref{tab:overall} summarizes the comprehensive benchmark results on the multi-class EEGmmidb dataset. Our GCMCG framework establishes new SOTA performance, substantially outperforming all competing methods across every metric. Critically, this superiority is most pronounced in the macro-averaged statistics that are vital for imbalanced datasets: GCMCG achieves a Macro-Rec of $79.41\%$ and a Macro-Prec of $83.07\%$, representing absolute improvements of $28.48\%$ and $24.94\%$ over the second-best overall accuracy DL competitor (FFCL), respectively. Together with a Macro-F1 score of $0.81$ (an increase of $0.27$ over FFCL) and a strong Cohen's Kappa coefficient of $0.82$, these results demonstrate that GCMCG maintains high per-class discriminability rather than being biased toward majority classes.

\begin{table}[htbp]
  \centering
  \caption{Comparison of the Top1-Acc across SOTA models on different datasets}
  \label{tab:datasets}
  \small % 或 \footnotesize
  \setlength{\tabcolsep}{5pt} % 适当压窄列间距
  \begin{tabular}{lcccccc} % 7 列：1 个 l + 6 个 c
    \toprule
    \textbf{Dataset} & FFCL & CMO-CNN & EEGNet& GRU-RNN & TD-LSTM  & \textbf{GCMCG} \\
    \midrule
    EEGmmidb    & 66.10  & 60.38 & 57.59 & 54.17 &  53.78 & \textbf{86.60} \\
    BCICIV-2A & 97.00 & 96.28 &  82.83 &  94.13& \textbf{99.57}  & 98.57 \\
    M3CV    & 92.71 & 91.93  &  87.43 & 91.48 &  85.10 & \textbf{99.61} \\
    \bottomrule
  \end{tabular}
\end{table}

To assess the generalization capability of GCMCG across heterogeneous EEG paradigms and recording conditions, we further evaluated Top1-Acc on two additional benchmark datasets: BCICIV-2a (motor imagery) and M3CV (multi-condition motor execution). Table~\ref{tab:datasets} presents this cross-dataset comparison against the most competitive DL baselines.

Our GCMCG framework exhibits exceptional robustness and consistency across all three benchmarks. It achieves the highest Top1-Acc on EEGmmidb ($86.60\%$) and M3CV ($99.61\%$), and remains highly competitive on BCICIV-2a with a Top1-Acc of $98.57\%$, only $1.00\%$ lower than the best-performing TD-LSTM ($99.57\%$). Notably, on M3CV GCMCG surpasses the second-best model (CMO-CNN, $91.93\%$) by a substantial $7.68\%$ absolute margin, highlighting its strong discriminative capability regardless of dataset complexity or recording protocol.

In stark contrast, competing methods display pronounced performance instability across datasets. TD-LSTM, while achieving the highest Top1-Acc on BCICIV-2a ($99.57\%$), drops to $53.78\%$ on EEGmmidb and $85.10\%$ on M3CV, revealing its sensitivity to domain shifts and noisier recording conditions. Similarly, FFCL's accuracy fluctuates from $97.00\%$ (BCICIV-2a) to $66.10\%$ (EEGmmidb), indicating limited scalability to more challenging, multi-class scenarios. EEGNet and GRU-RNN consistently underperform, with accuracies below $60\%$ on EEGmmidb and inferior results to GCMCG on all datasets, confirming their restricted representational capacity for diverse MI-ME decoding tasks. CMO-CNN shows relatively stable performance but remains inferior to GCMCG on two out of three datasets, underscoring the advantage of our proposed architecture.

Collectively, these results indicate that GCMCG achieves Top1-Acc that is competitive with or superior to existing methods across all three datasets---reaching state-of-the-art performance on EEGmmidb and M3CV and remaining very close to the best model on BCICIV-2a---highlighting its robustness and practical utility for motor brain--computer interfaces.

\subsection{Ablation Study}

To systematically evaluate the individual contributions of our framework, we conducted two complementary ablation studies: (1) the impact of the proposed denoising pipeline, and (2) the contribution of the core architectural components in the GCMCG model.

To assess the importance of denoising, we compared GCMCG on the raw EEGmmidb data and on its denoised counterpart. GCMCG shows a substantial performance increase from 76.68\% to 86.60\% on Top1-Acc, i.e., an absolute gain of 9.92\%. This confirms that denoising is not only beneficial but essential for fully exploiting the capacity of our framework. The 9.92\% absolute improvement directly validates the effectiveness of our denoising pipeline and helps explain the performance gap between our method and competing approaches that do not include this preprocessing stage.

\begin{figure*}[!t]
    \centering
    \begin{subfigure}[b]{0.48\textwidth}
        \centering
        \includegraphics[width=\textwidth]{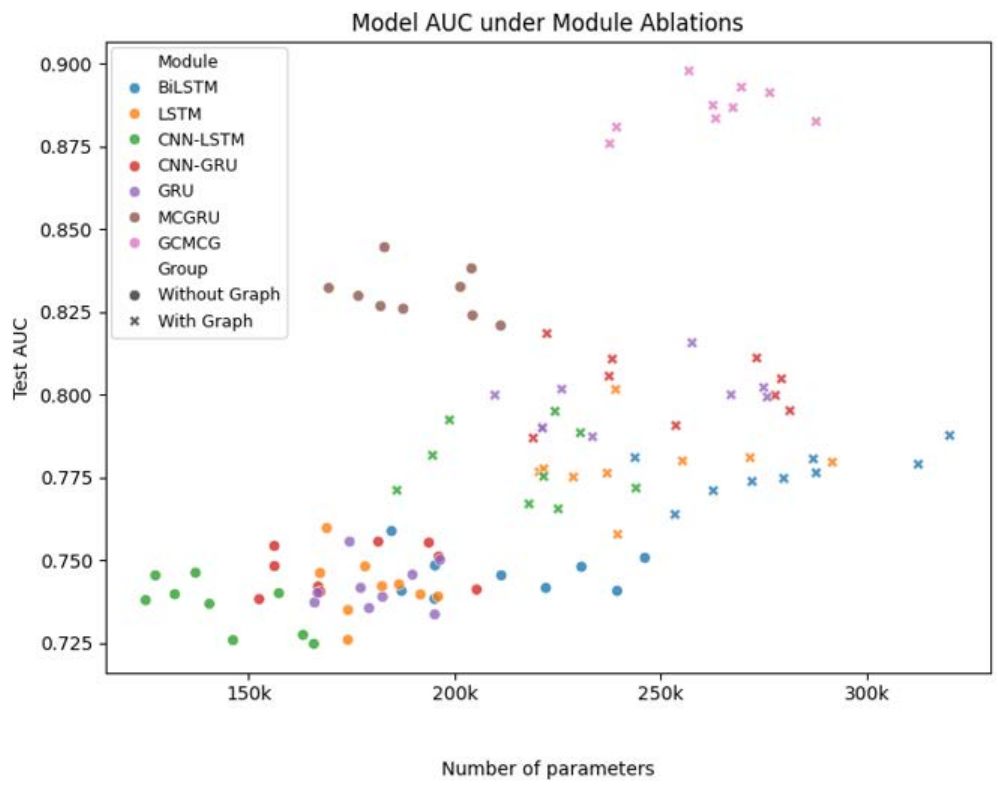}
        \caption{}\label{fig:ablation_a}
    \end{subfigure}
    \hfill
    \begin{subfigure}[b]{0.48\textwidth}
        \centering
        \includegraphics[width=\textwidth]{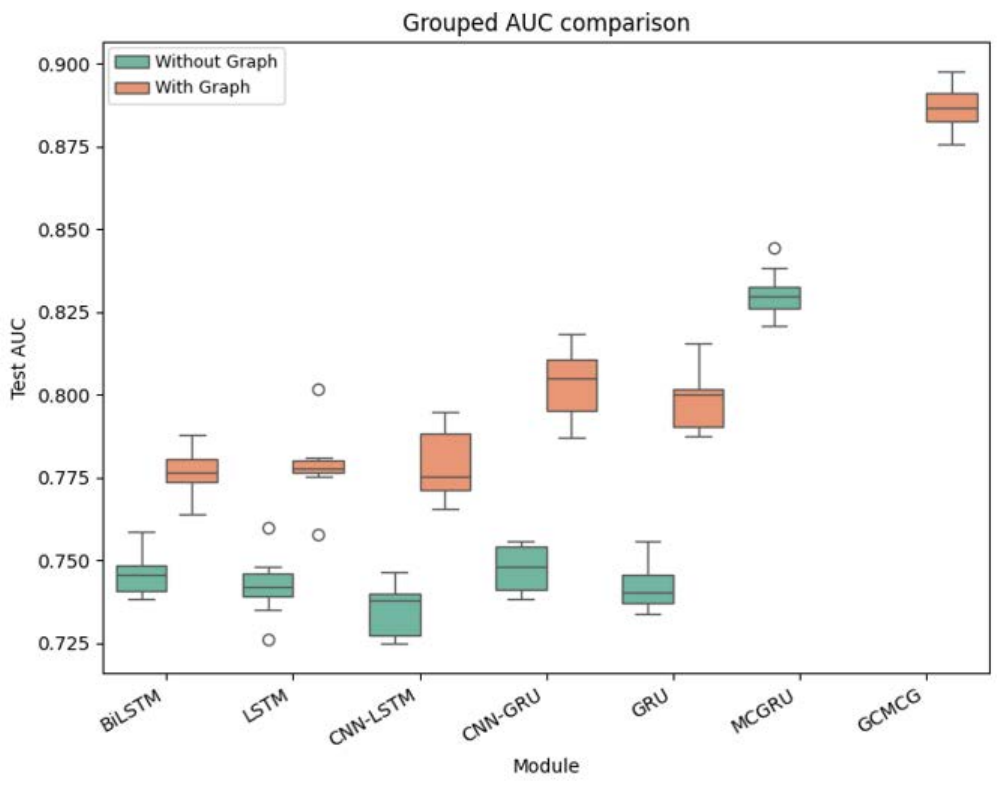}
        \caption{}\label{fig:ablation_b}
    \end{subfigure}
    
    \vspace{1ex}
    \caption{Performance comparison of GCMCG and its ablated variants on the EEGmmidb dataset. 
    The ``Module'' group represents different expert architectures. MCGRU denotes the Multi-Cluster GRU structure.
    The numbers in parentheses indicate the number of model parameters, which vary with the number of hidden layers and neurons. 
    \textbf{(a)} shows the classification performance (AUC) of each variant, while \textbf{(b)} presents the corresponding parameter counts.}\label{fig:ablation}
\end{figure*}

To better understand the contribution of each component in the GCMCG architecture, 
we then conduct a series of ablation experiments by selectively removing core modules—namely, 
the graph encoder and the clustering-based multi-expert module—and 
by replacing the underlying deep learning architecture used in the expert branches. 
All these ablations are evaluated under the same preprocessing, training, and evaluation conditions as the full model.
Notably, in the absence of the graph component, 
local experts are guided solely by clustering on EEG signals, 
without access to electrode topology or functional region priors.
This lack of structural context increases reliance on 
signal-based covariance matrices, 
resulting in greater computational complexity 
and potentially less coherent spatial grouping.
Figure~\ref{fig:ablation} summarizes the ablation results on the EEGmmidb dataset using two complementary visualizations. 
In subfigure~\subref{fig:ablation_a}, we present the classification performance of GCMCG and its ablated variants,
measured by the Area Under the ROC Curve (AUC). 
The variants use different expert backbones (LSTM, BiLSTM, CNN-LSTM, CNN-GRU) 
under comparable parameter settings to assess the impact of architectural choices.
Removing the graph module leads to the most noticeable drop in performance, decreasing test AUC by nearly 7\%, 
highlighting the critical role of graph-based spatial modeling. 
Eliminating the clustering module also degrades performance, confirming that function-specific electrode grouping enhances spatial specialization. 
The structure of the expert modules contributes meaningfully to performance differences. In particular, CNN-GRU achieves consistently better results 
than other architectures when model parameters are kept at similar levels, as shown in the left panel.
In subfigure~\subref{fig:ablation_b}, we display the corresponding model parameter counts for each variant. 
The numbers reflect adjustments in hidden layer depth and neuron number to balance the trade-off between model complexity and performance. 
The GCMCG model demonstrates its superiority not only in accuracy 
but also in computational efficiency relative to ablated variants with comparable or even higher parameter counts.
These results demonstrate that the combination of spatial attention, functional clustering, 
and carefully designed deep learning experts are all essential to GCMCG's superior performance in MI-ME EEG decoding.

\subsection{Feature visualization and interpretability}

To gain further insight into the behavior of modules, we visualize key components of GCMCG.
\begin{figure*}[!t]
    \centering
    \begin{subfigure}[b]{0.45\textwidth}
        \centering
        \includegraphics[width=\textwidth]{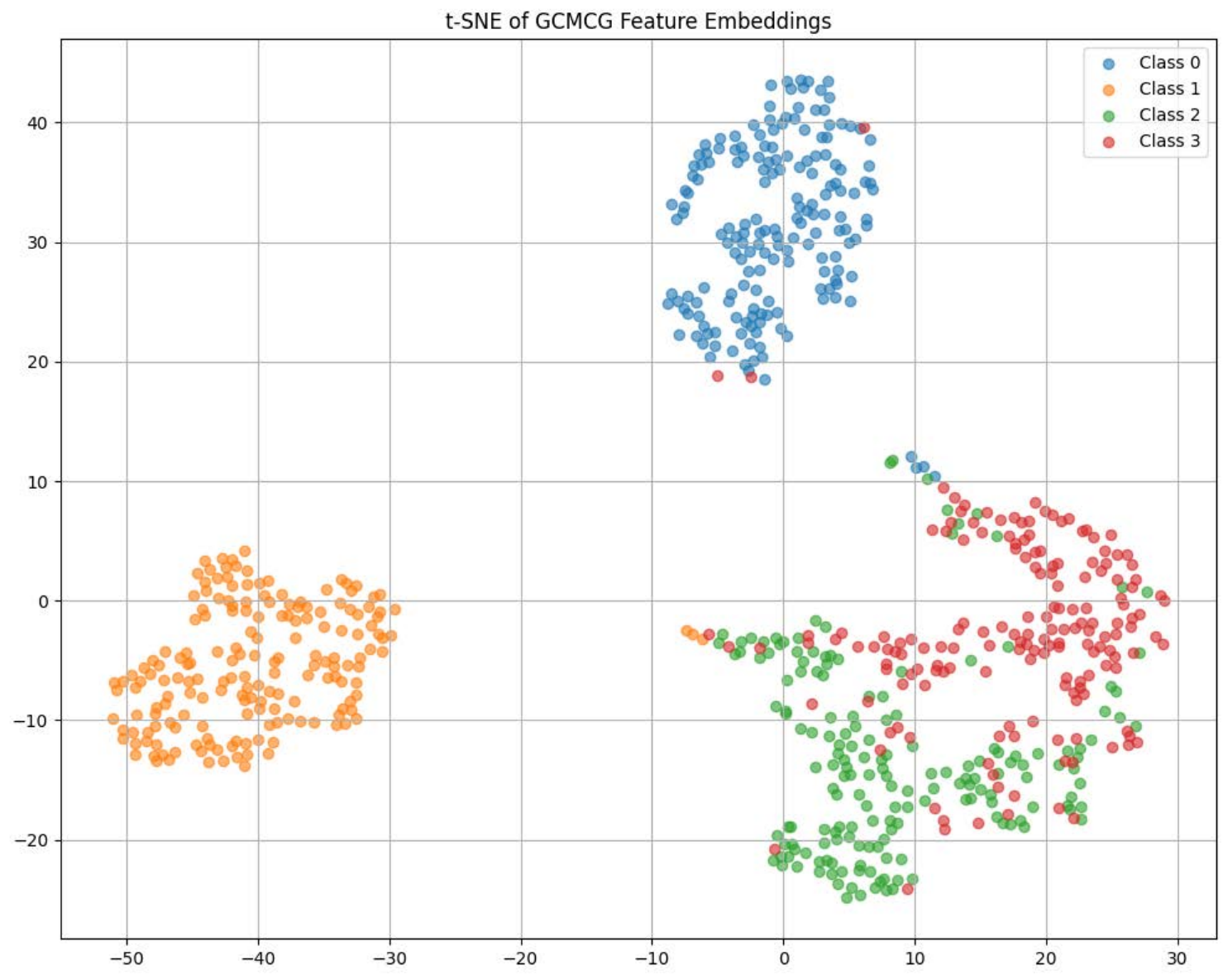}
        \caption{t-SNE plot of BCICIV-2A representations colored by class labels}\label{fig04_09:tsne-2a-2d}
    \end{subfigure}
    \begin{subfigure}[b]{0.45\textwidth}
        \centering
        \includegraphics[width=\textwidth]{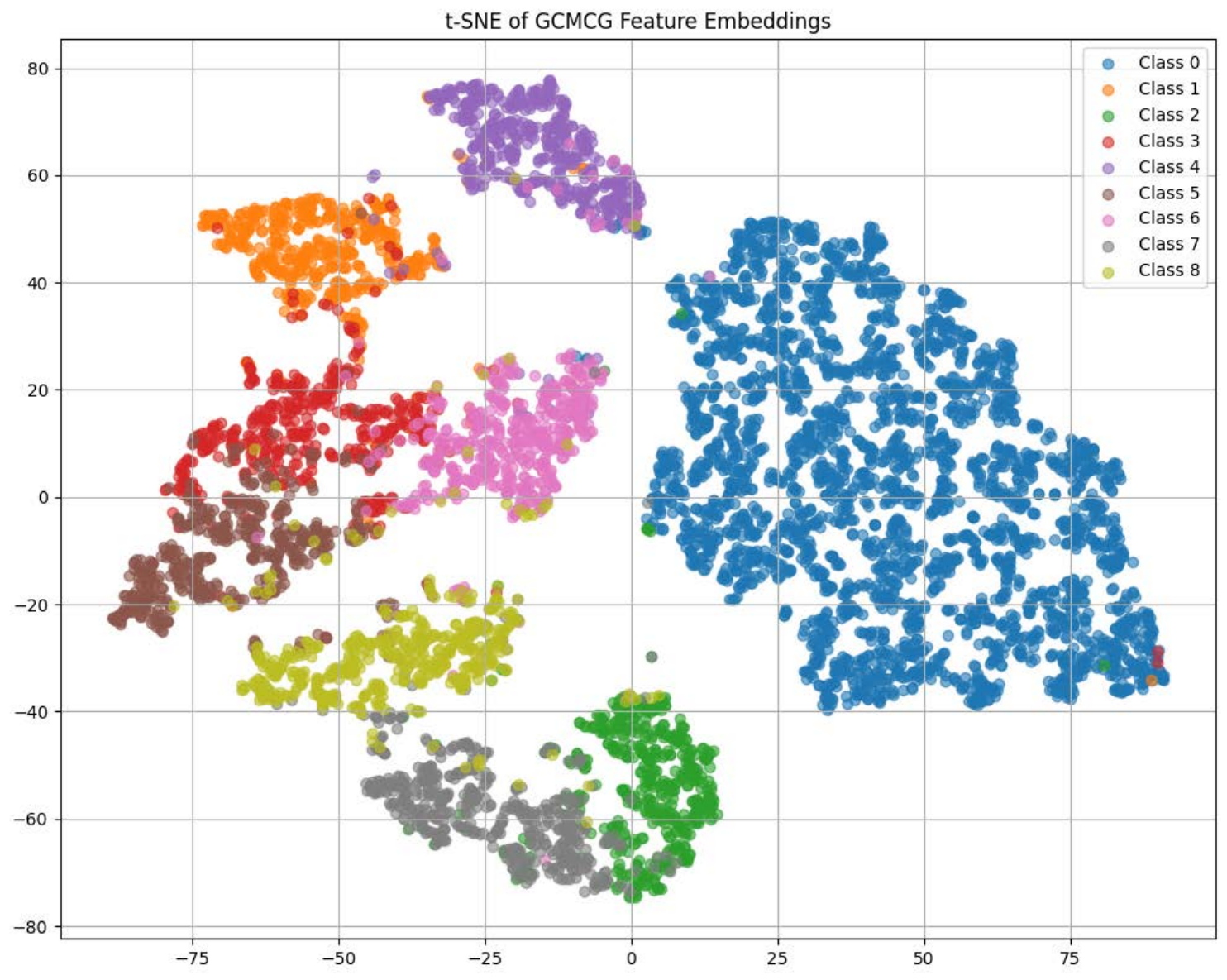}
        \caption{t-SNE 2D plot of EEGmmidb representations colored by class labels}\label{fig04_09:tsne-eegmmidb-2d}
    \end{subfigure}
    \hfill
    
    \vspace{1ex} % 留白
    \caption{t-SNE plot of final representations colored by class labels}\label{fig04_09:tsne}
\end{figure*}
\newpage  
As shown in Figure~\ref{fig04_09:tsne}, T-distributed stochastic neighbor embedding (t-SNE) shows 
the discriminative capacity of GCMCG's learned representations,  
and projects the final high-dimensional feature embeddings into a 2D space. 
Each point represents a test sample, and colors indicate ground-truth class labels.
Although some classes exhibit partial overlap—especially between neighboring motor tasks—the visualization shows 
that samples from the same class tend to form localized clusters. 
In particular, Class 0 (rest) forms a dense and well-separated region, 
suggesting that GCMCG effectively captures the distinction between motor-related and resting-state patterns. 
The mixed boundaries between motor execution and motor imagery classes also reflect 
the intrinsic signal similarities reported in prior studies.
The t-SNE projection qualitatively supports the effectiveness of GCMCG in learning structured, 
class-aware representations across multiple MI-ME paradigms, even under challenging cross-subject and multi-class conditions.

%\begin{figure}[htbp]
%    \centering
%    \includegraphics[width=0.65\linewidth]{}
%    \caption{Class-wise ROC and AUC curves for GCMCG}
%    \label{fig:roc}
%\end{figure}

%\begin{figure*}[!t]
%    \centering
    % include your combined tokenizer visual
%    \includegraphics[width=0.9\textwidth]{figures_xfz/tokenizer.png}
%    \caption{Visualization of the learned electrode tokenizer \(T_x\in\mathbb{R}^{C\times C}\) on the predefined eight‐neighborhood graph $\mathcal{G}$.  
%    {\bf(a)} Heat‐map of \(T_x[i,j]\).  
%    {\bf(b)} The same edges, colored by weight, overlaid on the 64‐channel scalp layout.}\label{fig04_11:tokenizer_vis}
%  \end{figure*}

  \begin{figure*}[!t]
    \centering
    \begin{subfigure}[b]{0.45\textwidth}
        \centering
        \includegraphics[width=\textwidth]{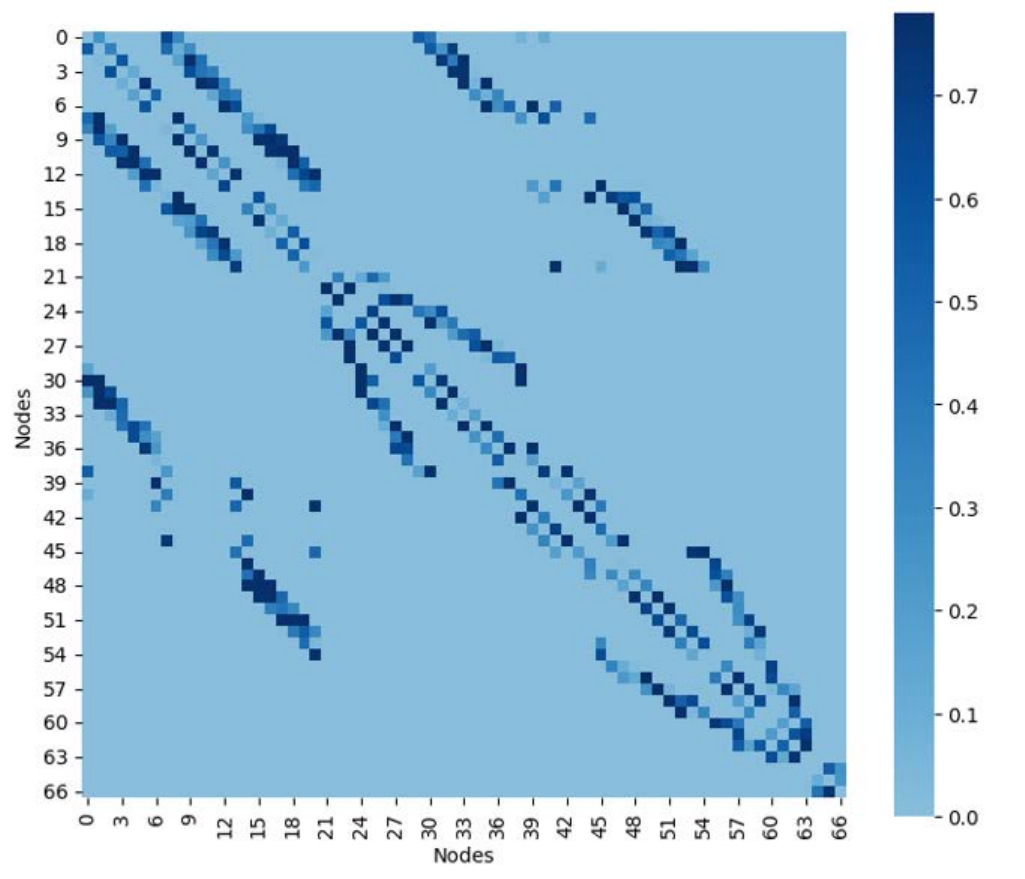}
        \caption{} % 建议删除空的caption 或者写“(a)”
        \label{fig04_11:tokenizer_vis_a}
    \end{subfigure}
    \hfill
    \begin{subfigure}[b]{0.45\textwidth}
        \centering
        \includegraphics[width=\textwidth]{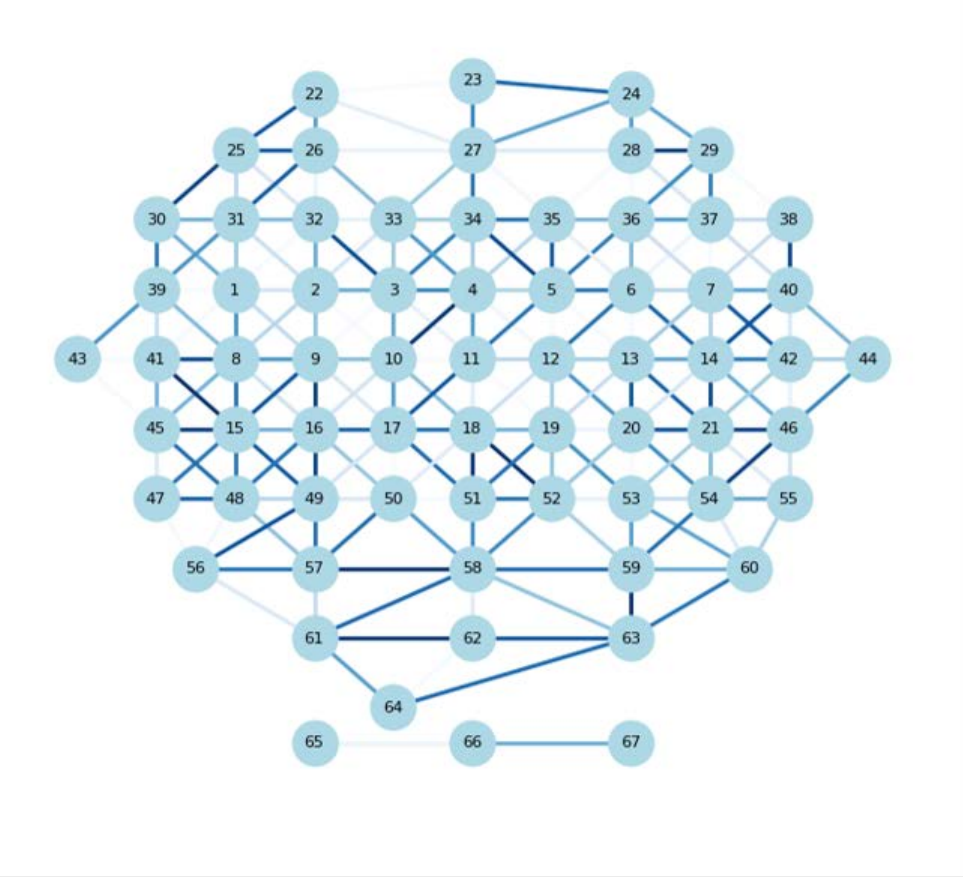}
        \caption{} % 建议删除空的caption 或者写“(b)”
        \label{fig04_11:tokenizer_vis_b}
    \end{subfigure}
    
    \vspace{1ex} % 留白
    \caption{Visualization of the learned electrode tokenizer \(T_x \in \mathbb{R}^{C \times C}\) on the predefined eight-neighborhood graph $\mathcal{G}$. 
    \textbf{(a)} shows the heat-map of pairwise weights, while \textbf{(b)} shows the anatomical projection of the same edges on the scalp layout.}\label{fig04_11:tokenizer_vis}
\end{figure*}

Figure~\ref{fig04_11:tokenizer_vis} illustrates 
how the trainable electrode tokenizer \(T_x\in\mathbb{R}^{C\times F}\) 
encodes spatial priors into edge weights along the eight-connected scalp graph.
In subfigure~\subref{fig04_11:tokenizer_vis_a}, the heat-map reveals strong, 
localized clusters of high weights around central motor channels (e.g., C3--CP3 and C4--CP4), 
reflecting the well-known event-related desynchronization 
in the sensorimotor rhythm~\cite{pfurtscheller1999event}. 
Intermediate couplings between homologous electrodes (e.g., left/right C3--C4) 
indicate learned cross-hemispheric coordination during both motor imagery and execution tasks. 
Peripheral channels (frontal and occipital sites) exhibit comparatively lower weights, 
consistent with their reduced involvement in MI/ME BCI paradigms.
Subfigure~\subref{fig04_11:tokenizer_vis_b} projects these learned edge weights onto 
the anatomical layout, with edges colored from light to dark blue according to their strength. 
The deepest blues concentrate in the central sulcus region, 
confirming that our GAT encoder strictly respects the manual adjacency list \(\mathcal{J}\) 
while adaptively emphasizing physiologically 
relevant pathways~\cite{rojas2018study}. 
This visualization validates the model's inductive bias 
and demonstrates its ability to discover meaningful functional connectivity beyond fixed, uniform weighting.
Importantly, the electrode tokenizer \(T_x\) is pre-trained on separate datasets  
(e.g., using BCI-IV 2a, M3CV when evaluating on EEGmmidb, and vice versa),
ensuring that the learned spatial priors generalize across recording montages and avoid information leakage. 
To further ensure the validity of our experimental results, 
the three datasets used in this study are explicitly partitioned into non-overlapping subsets: 
one for pre-training and the other for evaluation, without any subject or session overlap.
The GAT encoder subsequently auto-fits these fixed priors in a fully data-driven manner, 
learning the specific connectivity patterns inherent to the current dataset's electrode topology. 
This two-stage paradigm—generic spatial prior learning via tokenizer pre-training, 
followed by dataset-specific adaptation through GAT encoding—combines 
the strengths of transfer learning with on-the-fly graph refinement, 
achieving lower computational time and memory cost.

\subsection{Effectiveness of the Training Strategy}

\begin{figure*}[htbp]
    \centering
    \includegraphics[width=\linewidth]{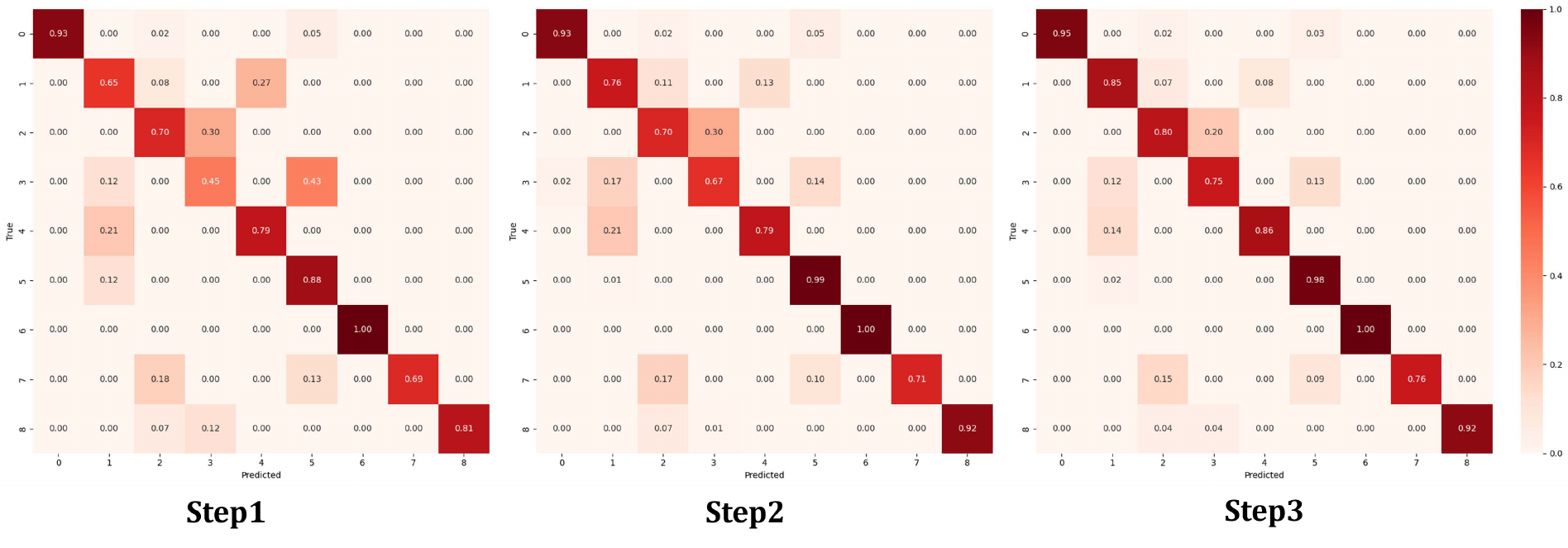}
    \caption{Confusion matrix for GCMCG on the EEGmmidb dataset}
    \label{fig:confmat}
\end{figure*}

Figure~\ref{fig:confmat} 
shows the normalized confusion matrix of GCMCG. 
As observed, the model distinguishes between imagery 
and execution tasks with high precision. 
Class-wise errors are primarily concentrated 
in adjacent motor classes such as ``imag\_both\_fist''
and ``move\_both\_fist''.

%\begin{figure*}[htbp]
%    \centering
%    \includegraphics[width=\linewidth]{figures_xfz/test_cm_tok.png}
%    \caption{Test Figure Confusion matrix for GCMCG on the EEGmmidb dataset}
%    \label{fig:confmat}
%\end{figure*}

\begin{figure}[htbp]
    \centering
    \includegraphics[width=0.8\linewidth]{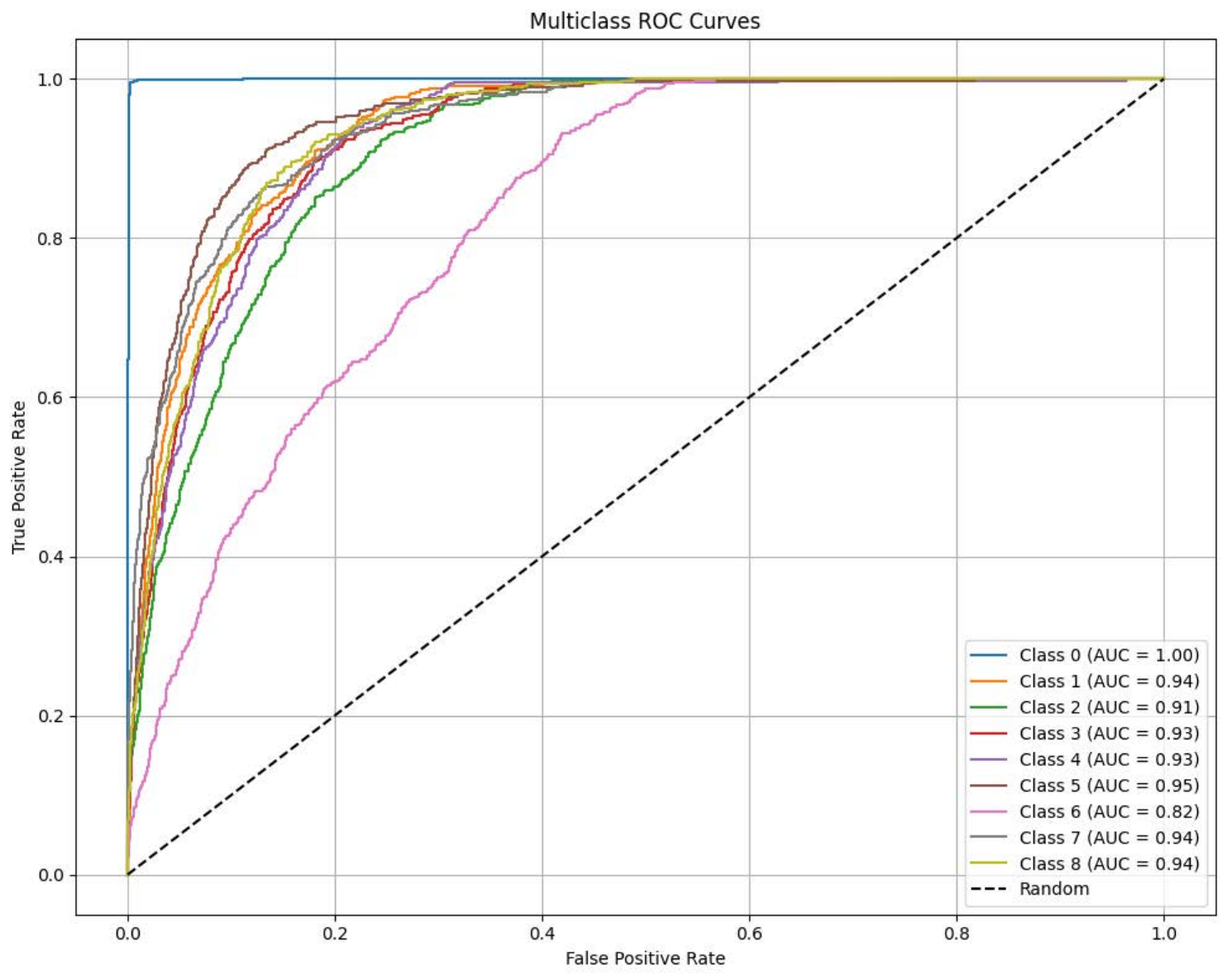}
    \caption{Class-wise ROC and AUC curves for GCMCG}
    \label{fig:roc}
\end{figure}

To evaluate discriminability across classes,  
we plot the class-wise Receiver Operating Characteristic (ROC) curves in Figure~\ref{fig:roc}. 
Each curve illustrates the trade-off between true positive rate (TPR) and false positive rate (FPR) for one of the nine classes, 
computed in a one-vs-rest manner.
As shown in the figure, GCMCG achieves consistently high Area Under Curve (AUC) scores across all classes, 
with an average AUC of 0.93. Notably, Class 0 (rest) attains a perfect AUC of 1.00, while most other classes exceed 0.90, 
indicating strong separability even under challenging multi-class conditions. The lowest score (0.82) is observed for Class 6, 
suggesting some overlap with neighboring classes, which is consistent with the t-SNE visualization.
The ROC curves confirm that GCMCG exhibits excellent discriminative ability and robustness across all MI-ME tasks, 
validating the effectiveness of its framework and training strategy.
Importantly, ROC analysis provides a comprehensive view of the classifier's behavior by illustrating the trade-off between Type I error (false positive rate) and Type II error (false negative rate).
This allows us to select appropriate decision thresholds to simultaneously control both error types, which is particularly advantageous when dealing with class imbalance. 
As a result, GCMCG maintains robust performance even in imbalanced scenarios, where traditional accuracy metrics may fail to reflect true discriminative power.

%\vspace*{\fill}

\section{Discussion}\label{sec:Discussion}

The importance of hybrid denoising is foundational to MI-ME decoding performance. The proposed two-stage denoising framework addresses a critical limitation of conventional pipelines: while notch and bandpass filters remove environmental artifacts, they cannot eliminate physiological noise due to spectral aliasing. Our ICA-WT stage specifically targets these artifacts by decomposing signals into independent components, applying wavelet thresholding, and reconstructing cleaned signals with maximally preserved neural information. This approach substantially enhances SNR, as validated by our ablation study where denoising yields a $9.92\%$ absolute improvement in Top1-accuracy for GCMCG. Notably, models lacking inherent denoising capabilities (e.g., TD-LSTM) show dramatic performance collapse on raw data, confirming that robust preprocessing is not merely complementary but essential for maximizing downstream spatial and temporal modeling efficacy. This hybrid strategy constitutes a distinct contribution that amplifies the framework's overall performance.

The inclusion of the graph encoder significantly enhances decoding performance, 
particularly under cross-subject conditions. 
This underscores the importance of incorporating spatial priors and topological 
context for robust generalization across varying electrode configurations. 
Unlike traditional architectures that treat EEG signals 
as Euclidean grids, 
the GCMCG framework leverages dynamic graph adaptation to more accurately reflect 
the functional organization of the brain. 
In practice, current non-embedded EEG acquisition systems 
often vary in electrode number and placement, 
complicating interpretation and limiting anatomical consistency across studies. 
GCMCG provides a generalizable solution to this issue 
by supporting graph encoding based on 
a customizable electrode adjacency list—currently supporting up to 67 nodes, 
and readily extensible to denser montages through an expanded adjacency schema.
Furthermore, the model bridges the gap between topological modeling,
medical EEG knowledge,  
and standard deep learning pipelines. 
Its tokenizer module, which can be pretrained independently, 
introduces a novel mechanism for encoding functional priors 
and anatomical layout into the learning process. 
This component not only facilitates spatial interpretability 
and cross-subject adaptability 
but also reduces training time and enhances model reuse—two challenges 
that persist in EEG research due to typically limited sample sizes 
and slow convergence. 
Importantly, the choice to avoid directly inputting EEG signals 
into the graph module improves computational efficiency 
and memory consumption, 
thereby enhancing the feasibility of real-time applications.

Functional clustering serves as a biologically inspired mechanism 
to segment the electrode space into interpretable subregions, 
each assigned to a specialized expert. 
This structure encourages localized temporal learning 
while mitigating interference from irrelevant channels. 
Empirical results confirm that region-specific modeling offers advantages 
over monolithic designs, particularly in complex motor decoding scenarios. 
Notably, our design allows seamless integration between clustering outputs 
and downstream deep learning modules, 
and supports the substitution of alternative clustering strategies, 
offering extensibility for broader use cases.

Model complexity, while often associated with performance, 
does not always correlate positively in EEG decoding tasks. 
Given the limited size of EEG datasets and the high inter-subject variability, 
overparameterization can lead to overfitting—especially 
when noise levels are high or task complexity is substantial. 
This aligns with observations in our experiments, 
where simpler expert configurations sometimes matched or outperformed larger ones.
Among the expert architectures evaluated, 
CNN-GRU demonstrates consistently better performance under comparable parameter constraints. 
This is consistent with previous findings that highlight GRU's favorable trade-off 
between expressive power and training efficiency. 
Compared to BiLSTM and CNN-LSTM, the CNN-GRU design enables faster convergence, 
better resistance to overfitting, and robustness in handling noisy temporal sequences, 
which are characteristic of non-invasive EEG signals.
Although model expressiveness generally benefits from increased architectural complexity, 
our findings suggest that well-structured designs—such as the use of gated fusion 
and selective expert routing—can achieve high accuracy without incurring significant parameter overhead. 
This is particularly valuable for practical BCI deployments, 
where model compactness and efficiency are critical.

The three-stage training strategy proposed in this work—comprising focal loss, 
progressively balanced sampling, and learnable weight scaling—substantially improves the model`s 
robustness to class imbalance and subject-level variability. 
In particular, the third stage, which involves freezing the backbone and tuning only the scaling factors, 
enables efficient adaptation with minimal overfitting risk. 
This modular training paradigm may prove beneficial for future BCI frameworks that require personalized fine-tuning.

\section{Conclusion and future work}

\label{sec:Conclusion}

This study proposed GCMCG, 
a generalizable EEG decoding framework that combines graph attention encoding, 
clustering-based regional specialization, and gated fusion of CNN-GRU experts. 
The model introduces a dynamic graph-based tokenizer to encode electrode topology, 
enabling adaptive spatial modeling. 
Spectral clustering is used to guide the dispatching of signals to localized GRU experts, 
while an entropy-regularized gating module adaptively fuses global and regional representations.
GCMCG achieves 86.60\%, 98.57\%, 
and 99.61\% classification Top1-accuracy on the EEGmmidb, BCI-IV 2a, and M3CV datasets,  
respectively, significantly outperforming previous models across multiple paradigms and class settings. 
The framework demonstrates strong robustness under class imbalance and cross-subject conditions, 
and supports variable-length sequences and flexible electrode layouts.
Future work includes freezing the tokenizer to reduce training cost, 
integrating domain generalization methods to improve cross-session transferability, 
and extending the framework to additional modalities such as EMG or eye-tracking. 
We also plan to explore dynamic clustering methods such as HMMs and refine modular components 
for real-time BCI applications with open-source deployment support.

While GCMCG demonstrates strong performance and generalization across complex MI-ME decoding tasks, 
several areas remain open for further exploration. 
One challenge lies in the robustness of spectral clustering when applied to low-SNR or highly nonstationary EEG segments. 
While the current pipeline infers functional regions via covariance-based clustering over graph-enhanced embeddings, 
adapting the clustering to dynamic or trial-specific contexts could further improve temporal stability and interpretability. 
Additionally, although the proposed tokenizer supports pretraining and generalization across datasets, 
its sensitivity to initialization and electrode configurations has not been fully explored. 
Investigating transfer learning strategies or incorporating subject-specific adaptation mechanisms—such as prototype alignment 
or manifold regularization—may help further extend the applicability of the model to heterogeneous populations and new paradigms without full retraining.

\section*{Acknowledgements}
% Thanks to \ldots
This work is partially supported by Basic Research Fund in Shenzhen Natural Science Foundation (No. JCYJ20240813104924033).


\section*{CRediT authorship contribution statement}
\textbf{Zijian Huang}: 
Conceptualization,
Methodology,
Investigation,
Supervision,
Project administration,
Funding acquisition.
\textbf{Yiqiao Chen}: 
Methodology,
Formal analysis,
Data curation,
Writing – Original Draft,
Visualization.
\textbf{Juchi He}: 
Validation,
Formal analysis,
Visualization.
\textbf{Fazheng Xu}: 
Methodology, 
Software,
Validation,
Formal analysis,
Resources,
Data curation,
Visualization,
Writing – Original Draft,

\appendix\label{appendix}
\section{Algorithms}\label{app1}

%%       \cite{<label>} ==> [1]

%%
%Example citation, See \cite{lamport94}.

\begin{algorithm}[htbp]
    \caption{Electrode Clustering via Spectral Method}\label{alg:03_02_clustering}
    \begin{algorithmic}[1]
    \REQUIRE{Node features $\mathbf{\Theta}' \in \mathbb{R}^{C \times D}$ from GAT output}
    \ENSURE{Cluster index map $\mathbf{k} \in {\{1,\ldots,K\}}^C$}
    
    \STATE{Compute pairwise correlation matrix $\mathbf{R} \in \mathbb{R}^{C \times C}$ from $\mathbf{\Theta}'$}
    \STATE{Construct graph Laplacian $\mathbf{L} = \mathbf{D} - \mathbf{R}$, where $\mathbf{D}$ is the degree matrix}
    \STATE{Perform eigendecomposition on $\mathbf{L}$ to get eigenvalues $\lambda_1 \leq \cdots \leq \lambda_C$}
    \STATE{Select optimal $K$ by maximizing eigengap: $K = \arg\max_i (\lambda_{i+1} - \lambda_i)$}
    \STATE{Form embedding $\mathbf{U} \in \mathbb{R}^{C \times K}$ from first $K$ eigenvectors}
    \STATE{Apply k-means clustering on $\mathbf{U}$ to obtain cluster labels $\mathbf{k}$}
    \RETURN{$\mathbf{k}$}
    \end{algorithmic}
    \end{algorithm}
    
%\begin{algorithm}[htbp]
%    \caption{Adaptive Expert Fusion with Entropy Regularization}\label{alg:03_03_fusion}
%    \begin{algorithmic}[1]
%    \REQUIRE{Expert features $\mathbf{v}_0, \mathbf{v}_1, \ldots, \mathbf{v}_K \in \mathbb{R}^{2D}$}
%    \ENSURE{Final fused feature $\hat{\mathbf{v}}_\text{fused} \in \mathbb{R}^{2D}$ and regularization loss $\mathcal{L}_\text{entropy}$}
%    
%    \STATE{Concatenate expert features into matrix: $\hat{\mathbf{v}} = [\mathbf{v}_0, \mathbf{v}_1, \ldots, \mathbf{Z}_K] \in \mathbb{R}^{(K+2) \times 2d}$}
%    \STATE Project features via linear layer: $\mathbf{q} = \mathbf{\tilde{F}}\mathbf{W}_{\text{gate}} + \mathbf{b}_{\text{gate}}$
%    \STATE Compute attention weights via softmax: $\mathbf{w_{\text{gate}}} = \text{Softmax}(\mathbf{q})$
%    \STATE Fuse features with weighted sum: $\mathbf{Z}_\text{fused} = \sum_{i=0}^{K+1} w_i \mathbf{Z}_i$
%    \STATE Compute entropy loss for sparsity: $\mathcal{L}_\text{gate} = -\sum_{i=0}^{K+1} w_i \log w_i$
%    \RETURN $\mathbf{Z}_\text{fused}, \mathcal{L}_\text{gate}$
%    \end{algorithmic}
%\end{algorithm}

\section{Equations}\label{app2}

\subsection{GRU}\label{equ1:gru}

\begin{equation}
    \begin{aligned}
        \mathbf{u}_t &= \sigma(\mathbf{W}_\text{update} \mathbf{z}_t + \mathbf{U}_\text{update} \mathbf{h}_{t-1}) \\
        \mathbf{r}_t &= \sigma(\mathbf{W}_\text{reset} \mathbf{z}_t + \mathbf{U}_\text{reset} \mathbf{h}_{t-1}) \\
    \end{aligned}
\end{equation}
    
Here,
$\mathbf{z}_t$ indicates the EEG signal after preprocessing, 
$\mathbf{u}_t$ denotes the update gate, 
$\mathbf{r}_t$ is the reset gate at time step $t$,
And $\mathbf{W}$ and $\mathbf{U}$ are weight matrices which are learned~\cite{cho2014learningGRU}. 
Both gates use the sigmoid function $\sigma(\cdot)$ to 
constrain their outputs between 0 and 1.
The update gate $\mathbf{u}_t$ determines how much of the previous hidden state $\mathbf{h}_{t-1}$ 
should be retained and carried forward, 
while the reset gate $\mathbf{r}_t$ controls how much of the past information to forget~\cite{dey2017gateGRU,luo2018rnnGRU}. 
The candidate activation $\tilde{\mathbf{h}}_t$ is computed as:
    
\begin{equation}
    \tilde{\mathbf{h}}_t = \tanh(\mathbf{W}_\text{hidden} \mathbf{z}_t + \mathbf{U}_\text{hidden} (\mathbf{r}_t \odot \mathbf{h}_{t-1}))
\end{equation}
    
The reset gate $\mathbf{r}_t$ modulates the contribution of 
the past hidden state when computing the new content. 
The final hidden state $\mathbf{h}_t$ is then derived as a linear interpolation between 
the previous hidden state and the candidate activation:
    
\begin{equation}
    \mathbf{h}_t = (1 - \mathbf{u}_t) \odot \mathbf{h}_{t-1} + \mathbf{u}_t \odot \tilde{\mathbf{h}}_t
\end{equation}
    
This formulation enables GRUs to effectively capture 
long-term dependencies with fewer parameters 
and lower computational cost compared to traditional LSTM units, 
offering a lightweight alternative that performs comparably 
in many EEG decoding scenarios, 
particularly under noise or resource constraints.

\section{Notations}\label{app3}

\begin{table*}[htbp]
    \captionsetup{labelfont=bf, textfont=normalfont, justification=raggedright, singlelinecheck=false}
    \caption{\\Learnable parameters in the proposed GCMCG framework.}\label{tab:03_01_params}
    \renewcommand{\arraystretch}{1.15}
    \begin{tabular*}{\linewidth}{@{\extracolsep{\fill}} lll>{\raggedright\arraybackslash}p{4cm}}
      \toprule
      \textit{Symbol} & \textit{Type} & \textit{Module / Function} & \textit{Description} \\
      \midrule
  
      \multicolumn{4}{l}{\text{Neural Network Weights}} \\
      $\mathbf{W}_{\mathrm{GAT}}$ & Matrix & Graph Attention Encoder & Weights for attention-based message passing \\
      $\mathbf{W}_{\mathrm{CNNGRU}}$ & Matrix & Global CNN-GRU & Spatial-temporal CNN-GRU parameters \\
      $\mathbf{W}_{\mathrm{MultiGRU}}^{(k)}$ & Matrix & Local Experts & Parameters for $k$-th local multi-GRU expert \\
      $\mathbf{W}_{\mathrm{gate}}$ & Matrix & Gating Network & Projection weights for expert fusion gate \\
      $\mathbf{b}_{\mathrm{gate}}$ & Vector & Gating Network & Bias term for expert fusion gate \\
      $\hat{\mathbf{W}}, \hat{\mathbf{b}}$ & Matrix / Vector & Classifier Head & Final classification weights and bias \\
  
      \midrule
      \multicolumn{4}{l}{\text{Graph-related Components}} \\
      $\mathbf{\Theta} $ & Matrix & Node features \\
      $\mathcal{G} = (\mathbf{\Theta}, \mathcal{J})$ & & & Electrode spatial graph with nodes and features \\
      $\mathbf{a}$ & Vector & Graph Encoder & Learnable attention vector for edge scoring \\

      \midrule
      \multicolumn{4}{l}{\text{Classifier and Regularization}} \\
      $\boldsymbol{\gamma}$ & Vector & LWS Head & Per-class scaling factors \\
      %$\lambda$ & Scalar & Regularization & Coefficient for entropy-based sparsity loss \\

      \bottomrule
    \end{tabular*}
  \end{table*}

  \begin{table*}[htbp]
    \centering
    \caption{Notation summary for non-learnable algebraic symbols used in Section~\ref{sec:Method}.}\label{tab:03_03_notations}
    \renewcommand{\arraystretch}{1.15}
    \begin{tabular*}{\linewidth}{@{\extracolsep{\fill}} lll}
    \toprule
    \textit{Symbol} & \textit{Type} & \textit{Description} \\
    \midrule
    \multicolumn{3}{l}{\text{General Inputs and Dimensions}} \\
    $C$ & Integer & Number of EEG channels (nodes) \\
    $S$ & Integer & Number of time steps per trial \\
    $D$ & Integer & Embedding dimension in hidden layers \\
    $K$ & Integer & Number of clusters  \\

    \midrule
    \multicolumn{3}{l}{\text{Graph Structure and Spectral Clustering}} \\
    $\mathcal{E}$ &  Dict & Customized electrode map \\
    $\mathcal{J}$ &  Dict / List &  Sparse adjacency list for graph construction \\
    $\mathcal{G} = (\mathcal{V}, \mathcal{E})$ & Graph & Electrode spatial graph with nodes and edges \\
    $\mathcal{G}_{\text{masked}}$ & Graph & Masked input node features used by GAT \\
    $\mathbf{L}$, $\mathbf{D}$ & Matrix & Graph Laplacian and degree matrix \\
    $\mathbf{U} \in \mathbb{R}^{C \times K}$ & Matrix & Spectral embedding from Laplacian eigenvectors \\
    $\lambda_i$ & Scalar & $i$-th eigenvalue of Laplacian \\

    \midrule
    \multicolumn{3}{l}{\text{Features and Embeddings}} \\
    
    $\mathbf{X} \in \mathbb{R}^{C \times S}$ & Matrix & Raw EEG signal matrix \\
    $\mathbf{Z}$ & Matrix & Standardized EEG signal  \\
    $\mathbf{\hat{v}} \in \mathbb{R}^{(K+2) \times 2D}$ & Matrix & Concatenated expert features \\

    \midrule
    \multicolumn{3}{l}{\text{Expert Gating and Output}} \\
    $\mathbf{g} \in \mathbb{R}^{K+2}$ & Vector & Gate logits from projection layer \\
    $\hat{\mathbf{y}}$ & Vector & Predicted class logits \\
    $y$ & Integer & Ground-truth class label \\
    
    \bottomrule
    \end{tabular*}
\end{table*}



\newpage
\bibliographystyle{elsarticle-num}  % 保留 Elsevier 数字引用样式

%\begin{thebibliography}{00}

%% For numbered reference style

%\bibitem{lamport94}
%  \textit{\LaTeX: a document preparation system},
%  Addison Wesley, Massachusetts,

%\end{thebibliography}
\end{document}